\documentclass[twocolumn]{aastex63}
\usepackage{color, colortbl}

\definecolor{Gray}{gray}{0.9}
\shorttitle{Spicule jets with thermal conduction}
\shortauthors{Gonz\'alez-Avil\'es et al.}

\begin{document}

\title{Spicule jets in the solar atmosphere modeled with resistive MHD and thermal conduction}

\correspondingauthor{Gonz\'alez-Avil\'es Jos\'e Juan}
\email{jjgonzalez@igeofisica.unam.mx}

\author[0000-0003-0150-9418]{J.J. Gonz\'alez-Avil\'es}
\affiliation{CONACyT - Servicio de Clima Espacial M\'exico - Laboratorio Nacional de Clima Espacial, SCiESMEX-LANCE. Morelia, Michoac\'an, M\'exico}
\affiliation{Instituto de Geof\'isica, Unidad Michoac\'an, Universidad Nacional Aut\'onoma de M\'exico. Morelia, Michoac\'an, M\'exico}

\author[0000-0002-1350-3673]{F.S. Guzm\'an}
\affiliation{Laboratorio de Inteligencia Artificial y Superc\'omputo. Instituto de F\'{\i}sica y Matem\'{a}ticas, Universidad Michoacana de San Nicol\'as de Hidalgo. Morelia, Michoac\'{a}n, M\'{e}xico.}

\author[0000-0002-0893-7346]{V. Fedun}
\affiliation{Department of Automatic Control and Systems Engineering, University of Sheffield, Sheffield, S1 3JD, UK}

\author[0000-0002-9546-2368]{G. Verth}
\affiliation{School of Mathematics and Statistics, University of Sheffield, Sheffield, S3 7RH, UK}

\begin{abstract}

Using numerical simulations, we study the effects of magnetic resistivity and thermal conductivity in the dynamics and properties of solar jets with characteristics of Type II spicules and cool coronal jets. The dynamic evolution of the jets is governed by the resistive MHD equations with thermal conduction along the magnetic field lines on a 2.5D slice. The magnetic field configuration consists of two symmetric neighboring loops with opposite polarity, used to support reconnection and followed by the plasma jet formation. In total 10 simulations were carried out with different values of resistivity and thermal conductivity, that produce jets with different morphological and thermal properties we quantify. We find that an increase in magnetic resistivity does not produce significant effects on the morphology, velocity and temperature of the jets. However, thermal conductivity affects both temperature and morphology of the jets. In particular, thermal conductivity causes jets to reach greater heights and increases the temperature of the jet-apex. Also, heat flux maps indicate the jet-apex and corona interchange energy more efficiently than the body of the jet. These results could potentially open a new avenue for plasma diagnostics in the Sun's atmosphere.
\end{abstract}


\section{Introduction}
\label{sec:Introduction}

Solar spicules are small-scale, jet-like plasma features observed ubiquitously in the solar chromosphere \citep{Beckers_1972,Sterling_2000,De_Pontieu_et_al_2004,De_Pontieu_et_al_2011}. Spicules may play an important role in energy and material supply to the upper layers of the solar atmosphere \citep{De_Pontieu_et_al_2011,Samanta_et_al_2019}. A number of theoretical models which related to formation and dynamics of spicules, including shocks wave plasma driving \citep{Sterling_2000,De_Pontieu_et_al_2004}, Alfv\'en waves \citep{Cranmer&Woolsey_2015,Iijima&Yokoyama_2017}, amplified magnetic tension \citep{Martinez-Sykora_et_al_2017a} or magnetic reconnection \citep{Ding_et_al_2011,Shelyag_et_al_2018}. Similarly, there are obsevations that highlight the evidence of magnetic reconnection in the generation of jets \citep{Borrero_et_al_2013,Martinez-Pillet_et_al_2011}. In particular, Type II spicules are collimated jets that reach maximum heights of 3-9 Mm and even longer in coronal holes, have a typical lifetime of 50-150 s \citep{De_Pontieu_et_al_2007a, Pereira_et_al_2012}, although these features can survive up to 500-800 s, while appearing and dissapearing in multiple chromospheric passbands, due to thermal evolution \citep{Pereira_et_al_2014}. Type II spicules show apparent upward motions with speeds of order 30-110 km s$^{-1}$ and temperatures of order $10^{4}$ K  \citep{Sterling_2000,Sterling_et_al_2010}. At the end of their life they usually exhibit rapid fading in chromospheric lines \citep{De_Pontieu_et_al_2007b,De_Pontieu_et_al_2017a}.

Numerical modelling is an important way for styling and analyzing various dynamical plasma processes in the solar atmosphere. In particular, it is a powerful tool for better understanding of transient phenomena such as jets.
For instance \citet{Takasao_et_al_2013}, study the acceleration mechanism of chromospheric jets associated with emerging fluxes using 2D MHD simulations. Similarly, there are more sophisticated models of jet formation in 2D and 3D, which have been performed by \citep{Isobe_et_al_2006,Pariat_et_al_2009,Archontis_et_al_2010, Jiang_et_al_2012_a}, that take into account physical effects such as magnetic resistivity and thermal conductivity. In particular, models including magnetic resistivity and thermal conductivity, are close to realistically describe the conditions in the solar atmosphere, as examples of this type of simulations we have \citep{Botha_et_al_2011}, where the authors use 3D MHD simulations to show that thermal conduction plays an essential role in the kink instability of coronal loops and cannot be ignored. Another example can be found in \cite{Fang_et_al_2014}, where the authors study the formation of coronal jets through the numerical simulation of the emergence of a twisted flux rope and found that field-aligned thermal conduction efficiently distributes the energy release, which is essential for comparing with synthetic emission. Apart from the ingredients of resistivity and thermal conductivity, there are other more sophisticated numerical simulations of Type II spicule formation that include the effect of radiation, partial ionization and ambipolar diffusion \citep{Martinez-Sykora_et_al_2009,Martinez-Sykora_et_al_2011,Martinez-Sykora_et_al_2017a,Martinez-Sykora_et_al_2017b,De_Pontieu_et_al_2017a}.

In this paper we continue the analysis carried out by \cite{2.5Dspicules}, by  including the thermal conductivity flux term in the resistive MHD equations, in addition to that we consider a wider range of resistivity values. In particular, we use a set of 4 realistic values of magnetic resistivity and thermal conductivity to analyze their effects on the morphology, maximum height, vertical velocity, thickness, temperature of the jet-apex and lifetime of the jets modelled. Aside of state of the art contribution in simulations, like partial ionization, radiation and ambipolar difussion such as in \citet{De_Pontieu_et_al_2017a,De_Pontieu_et_al_2017b,Nishizuka_et_al_2008,Nobrega-Siverio_et_al_2016,Martinez-Sykora_et_al_2012,Martinez-Sykora_et_al_2017a,Martinez-Sykora_et_al_2017b,Martinez-Sykora_et_al_2018,Yang_et_al_2013}, we show that thermal conductivity can modify the temperature, maximum height and width of the jets with some characteristics of Type II spicules and cool coronal jets. The carrying out the study in 2.5D allows us to control computing time in a flexible way and in the use of various parameter combinations to analyze the numerical simulations.

The paper is organized as follows. In Section \ref{Model&methods} we describe the resistive MHD equations with thermal conduction, the numerical methods we use, the model of solar atmosphere and the magnetic field configuration. In Section \ref{sec:results_numerical_simulations} we describe the parameters analyzed and the results of the numerical simulations for various experiments. Section \ref{sec:conclusions} contains conclusions and final comments.  

% ------------------------------------------
% ----->.    Section    <----------
% ------------------------------------------
\section{Model and methods}
\label{Model&methods}

% --------------------------------    
% ------> Subsection <------
% --------------------------------
\subsection{The system of resistive MHD equations with thermal conduction}

The model we consider to drive the plasma dynamics and the jet formation is the resistive MHD with thermal conduction and in particular we choose the Extended Generalized Lagrange Multiplier (EGLM) \citep{Jiang_et_al_2012_b} to be effective at keeping the evolution of jets under control \citep{2.5Dspicules,3dspicules}. The system of equations we use for the evolution of the plasma is given in \citet{Jiang_et_al_2012_b}, whose dimensionless version reads as follows:

\begin{eqnarray}
\frac{\partial\rho}{\partial t} +\nabla\cdot(\rho{\bf{v}})=0, \label{eglm_cont_equation} \\
\frac{\partial(\rho{\bf v})}{\partial t} + \nabla\cdot\left(\left(p+\frac{1}{2}{\bf B}^{2}\right){\bf I}+\rho{\bf vv}-{\bf BB}\right)\\\nonumber=-(\nabla\cdot{\bf B}){\bf B}+\rho{\bf g}, \label{eglm_mom_equation} \\
\frac{\partial E}{\partial t}+\nabla\cdot\left({\bf v}\left(E+\frac{1}{2}{\bf B}^{2}+p \right)-{\bf B}({\bf B}\cdot{\bf v})\right)\nonumber \\=-{\bf B}\cdot(\nabla\psi)-\eta\nabla\cdot({\bf J}\times{\bf B})+ \nabla\cdot {\bf q} + \rho{\bf g}\cdot{\bf v}, \label{eglm_energy_equation} \\[0.3 cm]
\frac{\partial{\bf B}}{\partial t}+\nabla\cdot({\bf Bv}-{\bf vB}+\psi{\bf I})=-\eta\nabla\times{\bf J}, \label{eglm_induction_equation} \\
\frac{\partial\psi}{\partial t}+c_{h}^{2}\nabla\cdot{\bf B}=-\frac{c_{h}^{2}}{c_{p}^{2}}\psi, \label{eglm_psi_equation} \\ \nonumber
{\bf J}=\nabla\times{\bf B}, \\ \nonumber
E=\frac{p}{(\gamma-1)}+\frac{\rho{\bf v}^{2}}{2}+\frac{{\bf B}^{2}}{2}, \\ \nonumber
\end{eqnarray}

\noindent where $\rho$ is the mass density, ${\bf v}$ is the velocity field, ${\bf B}$ is the magnetic field, $E$ is the total energy density, the plasma pressure $p$ is described by the equation of state of an ideal gas $p=(\gamma -1)\rho e$, where $e$ is the internal energy and $\gamma$ its adiabatic index, ${\bf g}$ is the gravitational field at the solar surface, ${\bf J}$ is the current density, $\eta$ is the magnetic resistivity and $\psi$ is a scalar potential that helps damping out the violation of the constraint $\nabla\cdot{\bf B}=0$. Here $c_h$ is a wave speed and $c_p$ is the damping rate of the wave of the characteristic mode associated to $\psi$. In our simulations we use $c_p=\sqrt{c_r}c_h$, with $c_r=$0.18 and $c_h=$0.001, that have shown useful for 2.5D and 3D simulations.

The contribution of thermal conduction is included in the equation for the energy (\ref{eglm_energy_equation}), through the heat flux vector that allows the heat propagation along the magnetic field lines \citep[see e.g.][]{Jiang_et_al_2012_a}

\begin{equation}
{\bf q} = \kappa T^{5/2}({\bf B}\cdot\nabla T){\bf B}/B^{2}, \label{q_along_field_lines}
\end{equation}

\noindent where $\kappa$ is the thermal conductivity of the plasma, $T$ its temperature and ${\bf B}$ is the magnetic  field. 

% ---------------------------------------------------
% ---------->    SUB-SECTION    <----------
% ---------------------------------------------------

\subsection{Numerical methods}

The evolution of jets is analyzed with a 2.5D approach, which consists in the restriction of the dynamics under consideration to have slab symmetry along one of the two horizontal directions of the spatial domain. We use Cartesian coordinates $x$ and $z$ describing the two-dimensional domain, and $y$ is the direction of the slab symmetry.

We solve numerically the resistive EGLM-MHD equations with thermal conduction given by the system of equations (\ref{eglm_cont_equation})-(\ref{eglm_psi_equation}) using the Newtonian CAFE code \citep[see e.g.][]{NewtonianCAFE,CAFE-Q}, on a uniform cell centered grid, using the method of lines with a third order total variation diminishing Runge-Kutta time integrator described in \citet{Shu&Osher_1989}. In order to use the method of lines, the right hand side of equations (\ref{eglm_cont_equation})-(\ref{eglm_psi_equation}) are discretized using a finite volume approximation with High Resolution Shock Capturing methods, e.g., \citet{LeVeque_1992}. For this, we first reconstruct the variables at cell interfaces using the minmod limiter, and numerical fluxes are calculated using the Harten-Lax-van Leer-Contact (HLLC) approximate Riemann solver \citep[see e.g.][]{Li_2005}. 

% ---------------------------------------------------
% ---------->    SUB-SECTION    <----------
% ---------------------------------------------------

\subsection{Model of the solar atmosphere and magnetic field configuration}

\begin{figure*}
\centering
\includegraphics[width=7.0cm,height=6.0cm]{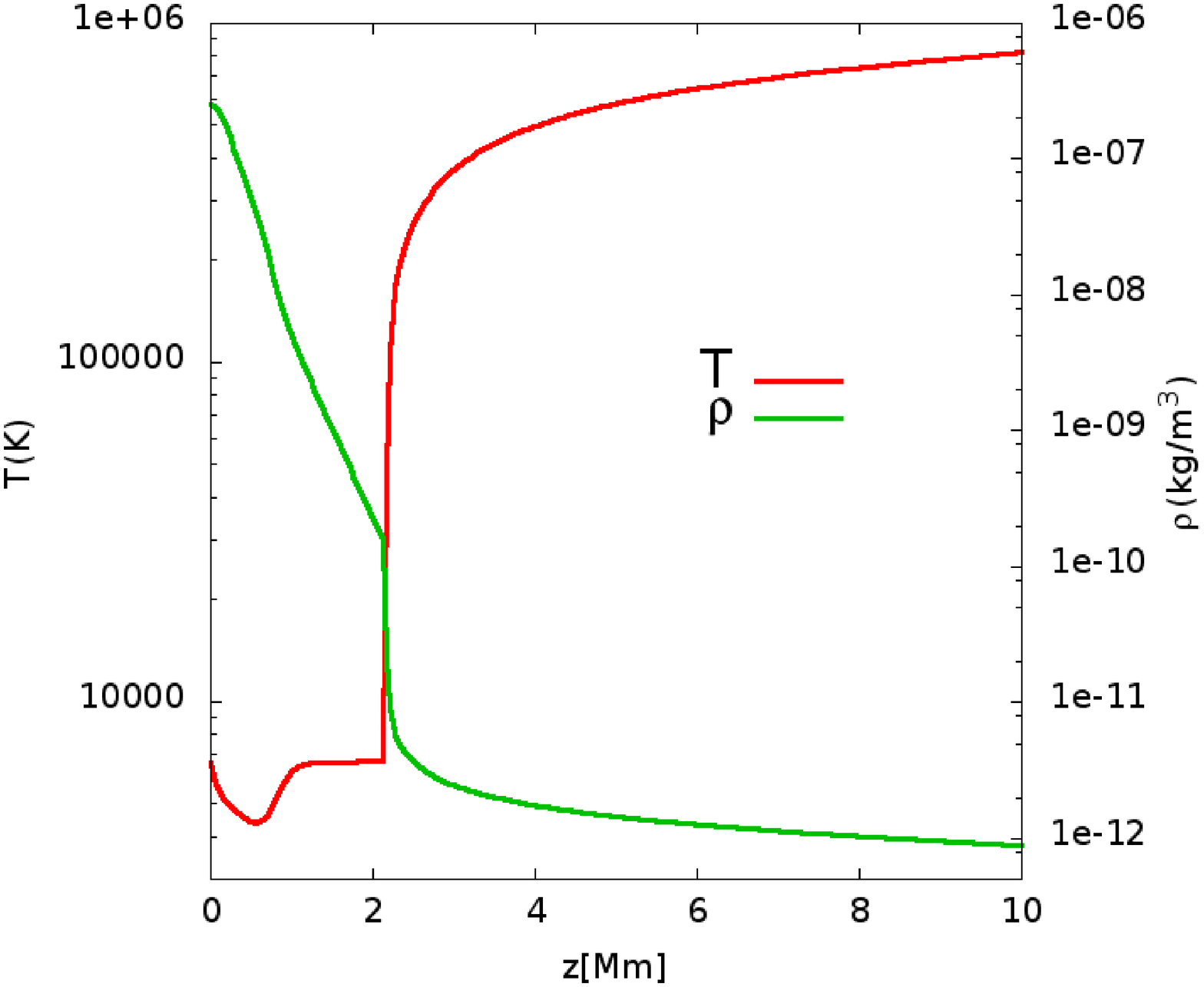}\\
\includegraphics[width=12.5cm,height=4.5cm]{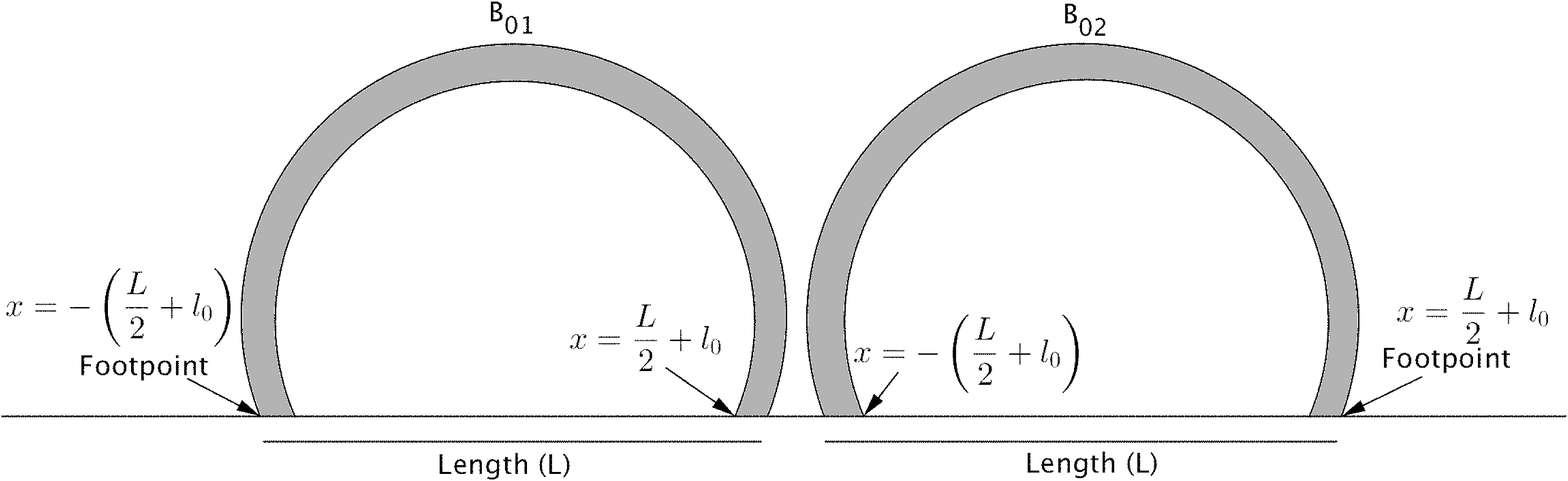} 
\caption{Top: Temperature (red) and mass density (green) as a function of height for the C7 equilibrium solar atmosphere model. Bottom: two consecutive symmetric magnetic loop configurations with the same field strength $B_{01}=B_{02}=B_{0}$. These pictures were taken from \cite{2.5Dspicules}.}
\label{fig:1}
\end{figure*}

At initial time of simulation, we assume the solar atmosphere to be in hydrostatic equilibrium, with the temperature field consistent with the semiempirical C7 model of the chromosphere transition region \citep{AvretLoeser2008}. The density and temperature profiles used to start the simulations are shown at the top of Figure \ref{fig:1}. For the equation of state we assume an adiabatic index $\gamma=5/3$ and the gravitational field is set to ${\bf g}=-g\hat{z}$ with $g=274$m s$^{-2}$ in the equations of momentum and energy, more details can be found in \cite{2.5Dspicules}.

The magnetic field configuration is a superposition of two neighboring loops and is constructed from a potential. Based on \cite{Priest_1982,DelZannaetal2005} and \cite{2.5Dspicules}, the magnetic field potential for two symmetric loops, that decreases with height exponentially, is given by

\begin{equation}
A_y(x,z) = \frac{B_{0}}{k}(\cos(k(x + l_0))+\cos(k(x - l_0)))\exp(-kz) , \label{vector_potential_translation}
\end{equation}

\noindent where $l_0$ parametrizes the position of the foot points for each of the loops and $B_{0}$ is the magnetic field strength. In this paper, we use the parameters $B_0=40$ G and $l_0=3.5$ Mm, because they are physically sound and produce jets with the some properties of Type II spicules successfully \citep{2.5Dspicules}. In this manner we concentrate on the properties of the ejected structure. A schematic picture of the magnetic configuration is shown to the bottom of Figure \ref{fig:1}. 

Further details of the set up are that we fix $k=\pi/L$ with $L=8$ Mm, where $L$ is the distance between the two footpoints of the loop. Then we execute simulations in the 2D domain $x\in[-4,4] \times z\in[0,10]$ in units of Mm, covered with 300$\times$375 grid cells in $x$ and $z$ direction correspondingly. Since we are using a three dimensional code, we cover the additional direction $y$ with four cells. The boundary conditions used are outflow at all faces of the domain. 

% -------------------------------------------
% ----->     SECTION     <-----
% -------------------------------------------
\section{Results of numerical simulations and analysis}
\label{sec:results_numerical_simulations}

% -------------------------------------------
% ----->     SUB-SECTION     <-----
% -------------------------------------------

\subsection{Parameters analyzed}

We study the effects of magnetic resistivity and thermal conductivity on jet formation process with characteristics of Type II spicules. The following parameters were analyzed in details for different runs: the maximum height, width of jet-apex, average temperature of the jet-apex at maximum height and time when the jet reaches the maximum height obtained with the various combinations of the resistivity $\eta$ and thermal conductivity $\kappa$. 

With the various combinations of these parameter values we define 10 cases specified in Table \ref{table1}, the combinations were chosen taking into account the level of realism of the values. For ease, in what follows, we will use the code unit values of resistivity and thermal conductivity to define and describe the various scenarios in our analysis.

For resistivity we use the values $\eta =5\times 10^{-2},~1\times 10^{-1},~2\times 10^{-1}, ~3\times 10^{-1} \Omega\cdot \mathrm{m}$, which are reasonable values for a fully ionized solar atmosphere \citep{Priest_2014}. For thermal conductivity we use the values $\kappa=0,~7.6\times 10^{-13},~10^{-12},~9\times 10^{-12},~10^{-11}$ W m$^{-1}$ K$^{-7/2}$, where $10^{-11}$ W m$^{-1}$ K$^{-7/2}$ is a typical value in the chromosphere and $9\times 10^{-12}$ W m$^{-1}$ K$^{-7/2}$ is a typical value in the corona for a fully ionized gas \citep{Spitzer_1962,Botha_et_al_2011}. We consider the fully ionized condition is acceptable due to the time scales of the jet processes obtained in this paper. For example, at the chromospheric level, the time-scale of neutron-ion collision frequency lies between $1-10^4$ Hz \citep[see e.g. Figure 4 of][]{Martinez-Sykora_et_al_2012}, which is a time scale at least one order of magnitude smaller than that of the jet evolution shown in our simulations. With respect to thermal conductivity, we use values that range from those suitable for the chromosphere for a partially ionized plasma, up to those appropriate for the fully ionized plasma at the corona \citep[see e.g.][]{Spitzer_1962,Botha_et_al_2011}. Following the conventions of \cite{Gonzalez-Aviles&Guzman_2015} to make the MHD equations dimensionless, we obtain the dimensionless values of $\bar{\eta}$ and $\bar{\kappa}$ according to the following scaling:

\begin{eqnarray}
\bar{\eta} = \frac{\eta}{L_{0}\mu_{0}v_{0}}, \\
\bar{\kappa} = \frac{\kappa T_{0}^{7/2}}{L_{0}\rho_{0}v_{0}^{3}},
\end{eqnarray}

\noindent where $L_{0}=10^{6}m$, $\mu_{0}=4\pi\times10^{-7}$, $v_{0}=10^{6} \mathrm{m}/s$, $\rho_{0}=1\times10^{-12} kg/\mathrm{m}^{3}$ and $T_{0}=$7.269$\times10^{7} K$. Therefore the values in code units for resistivity and thermal conductivity constants are $\eta=3.97\times10^{-8},~8\times10^{-8},~ 1.6\times10^{-7},~ 2.5\times10^{-7}$ and $\kappa=0,~2500,~3275,~29479,~32755$. 

%----- Table 1 -----%

\begin{deluxetable*}{cccccccc}
\tablenum{1}
\tablecaption{Values of $\eta$ and $\kappa$ used in our study. For each case we list the following properties of the simulated jet: maximum height, width of the jet-apex at $h_{max}$, average of temperature of the jet-apex at $h_{max}$ and time when the jet reaches the maximum height\label{table1}.}
\tablewidth{0pt}
\tablehead{
\colhead{Run \#} & \colhead{$\eta$} & \colhead{$\kappa$} & \colhead{$h_{max}$ (Mm)} & \colhead{width (Mm)} & \colhead{$T_{head}$ (K)} & \colhead{$t_{hmax}$ (s)}
}
\startdata
\rowcolor{Gray}
1 & 3.97e-8 & 0 & 7.3 & 1.1  & 56240 & 210 
\\
2 & 8e-8 & 0 & 7.3 & 1.1 & 57269 & 210  
\\
3 & 1.6e-7 & 0 & 7.3 &  1.1 & 56678 & 210 
\\
4 & 2.5e-7 & 0 & 7.3 & 1.1  &  58643 & 210 
\\
5 & 3.97e-8 & 2500 &  7.5  & 0.92  & 62485  & 210 
\\
\rowcolor{Gray}
6 & 3.97e-8 & 3275 & 7.5  & 0.92  & 63920 & 210 
\\
7 & 3.97e-8 & 29479 & 7.7  & 0.82 & 69118 & 210 
\\
8 & 3.97e-8 & 32755 & 7.7 & 0.82  & 73049 & 210 
\\
9 & 8e-8 & 2500 & 7.5 & 0.92 & 55395   & 210 
\\
\rowcolor{Gray}
10 & 8e-8 & 32755 & 7.7 &  0.82 & 61714 & 210 
\\
\enddata
\end{deluxetable*}

% -------------------------------------------
% ----->     SUB-SECTION     <-----
% -------------------------------------------

\subsection{Results of numerical simulations}

\begin{figure*}
\centering
\centerline{\Large \bf   
      \hspace{0.16 \textwidth}  \color{black}{\normalsize{(a)}}
      \hspace{0.2\textwidth}  \color{black}{\normalsize{(b)}}
      \hspace{0.19\textwidth}  \color{black}{\normalsize{(c)}}
         \hfill}
\includegraphics[width=4.3cm,height=5.5cm]{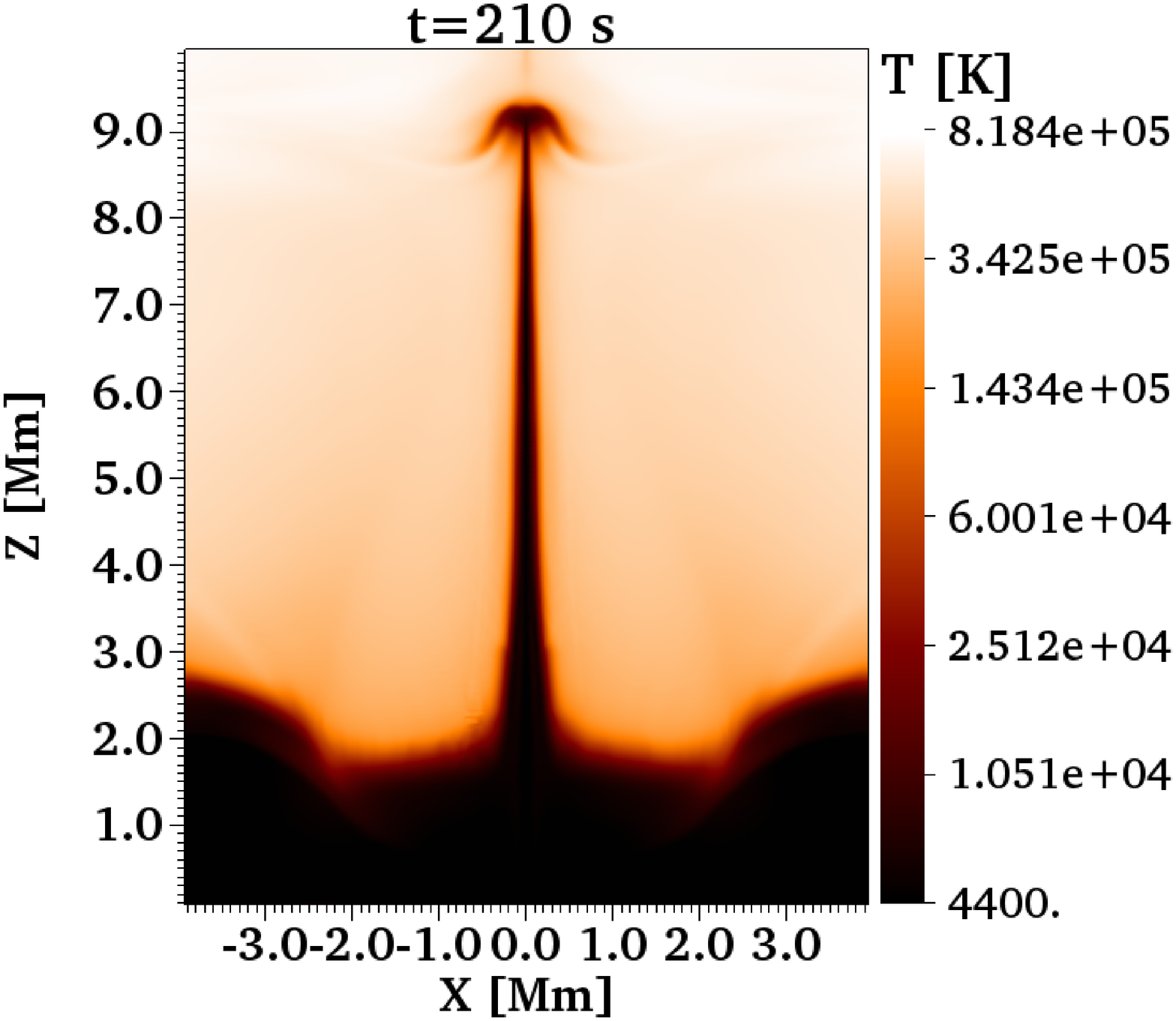}
\includegraphics[width=4.3cm,height=5.5cm]{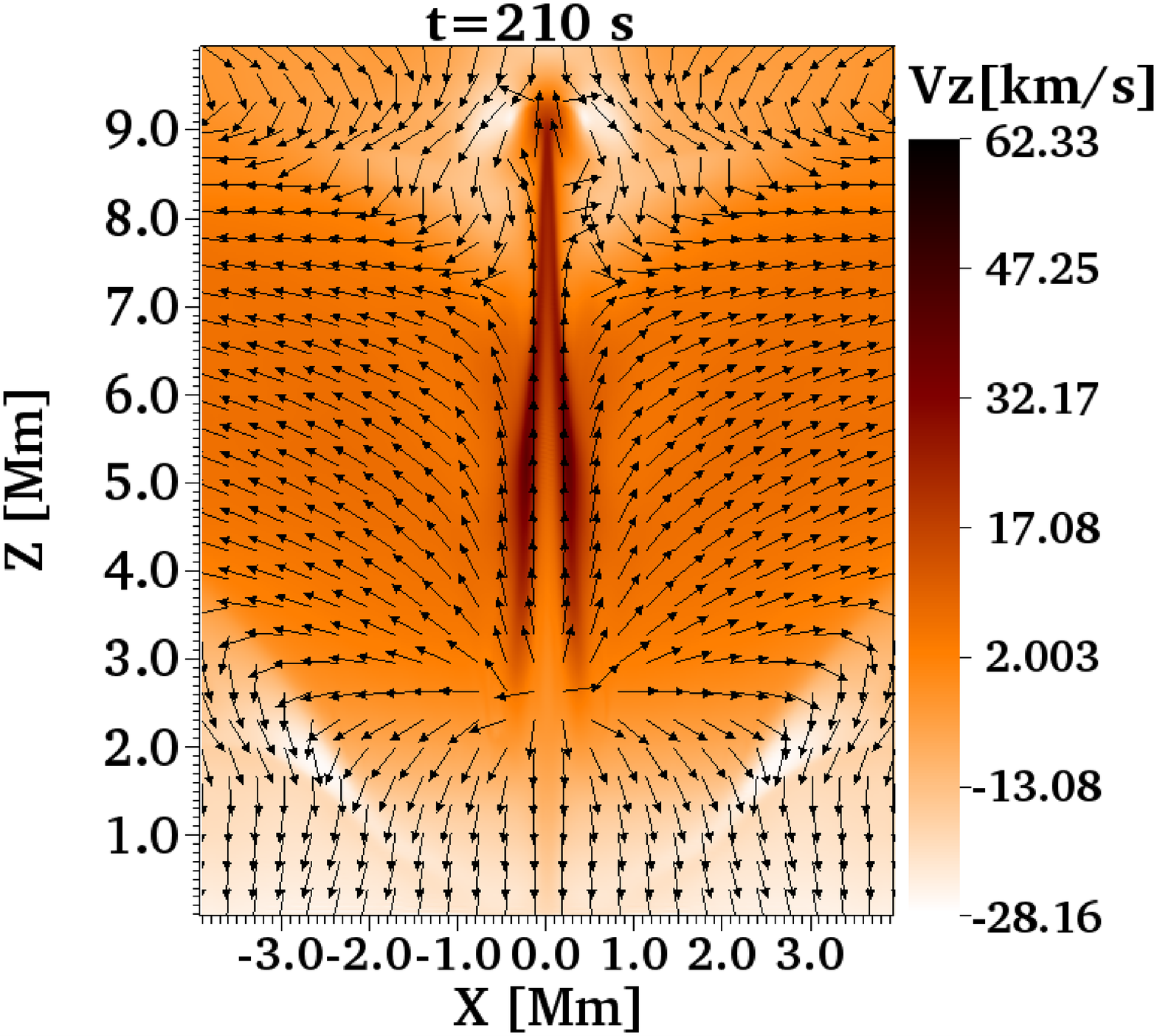} 
\includegraphics[width=4.3cm,height=5.5cm]{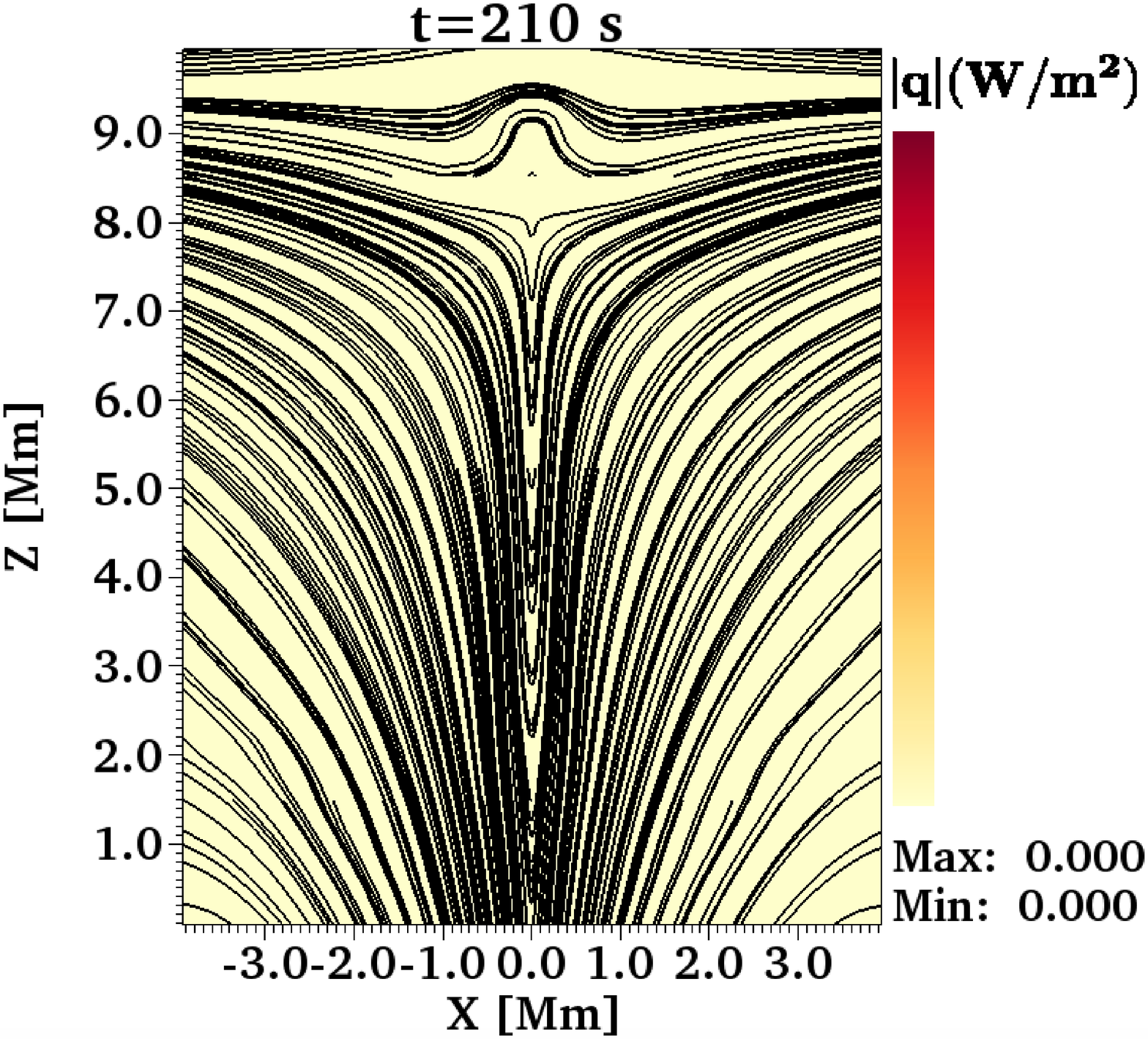} \\
\centerline{\Large \bf   
      \hspace{0.16 \textwidth}  \color{black}{\normalsize{(d)}}
      \hspace{0.2\textwidth}  \color{black}{\normalsize{(e)}}
      \hspace{0.19\textwidth}  \color{black}{\normalsize{(f)}}
         \hfill}
\includegraphics[width=4.3cm,height=5.5cm]{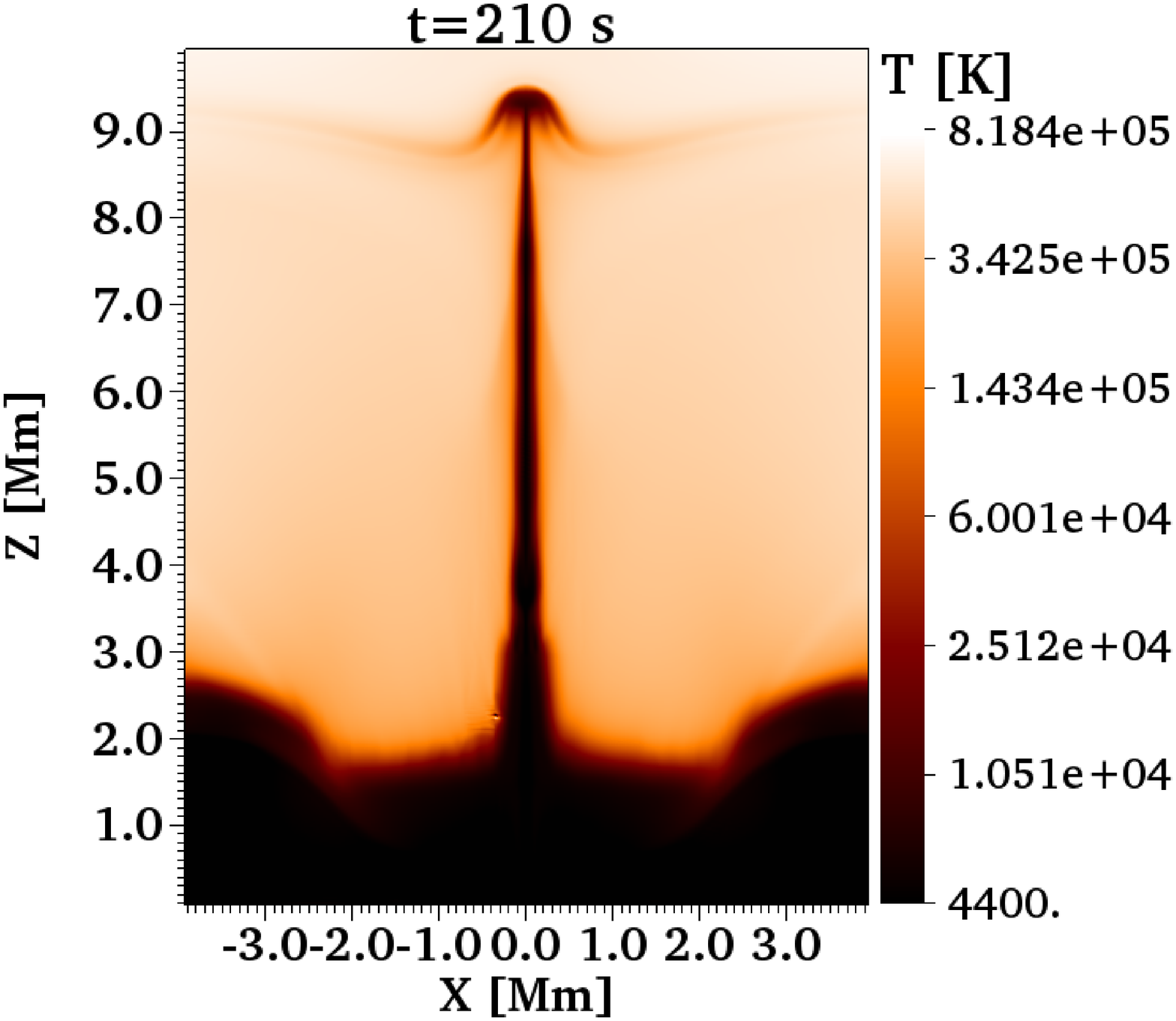}
\includegraphics[width=4.3cm,height=5.5cm]{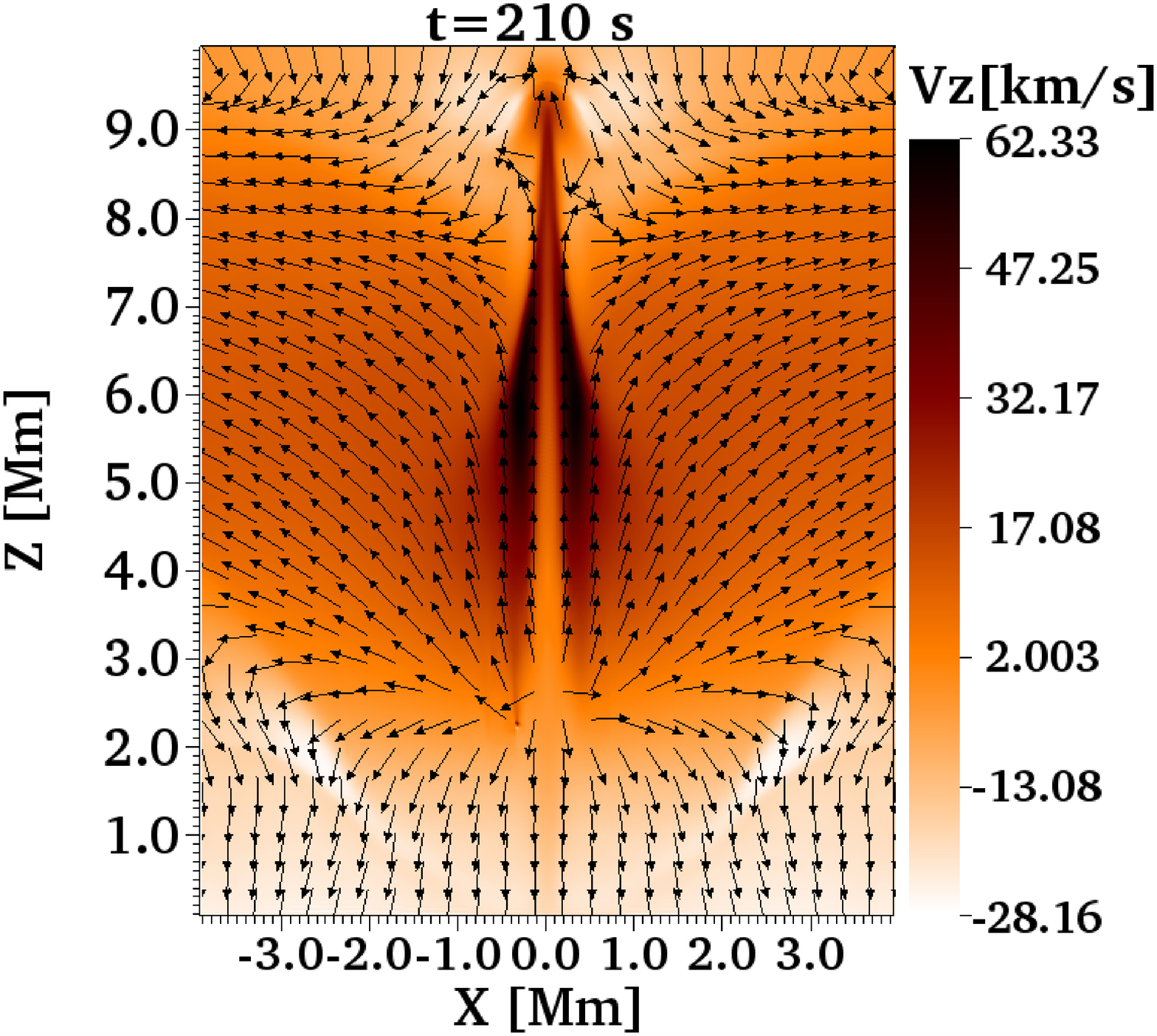}
\includegraphics[width=4.3cm,height=5.5cm]{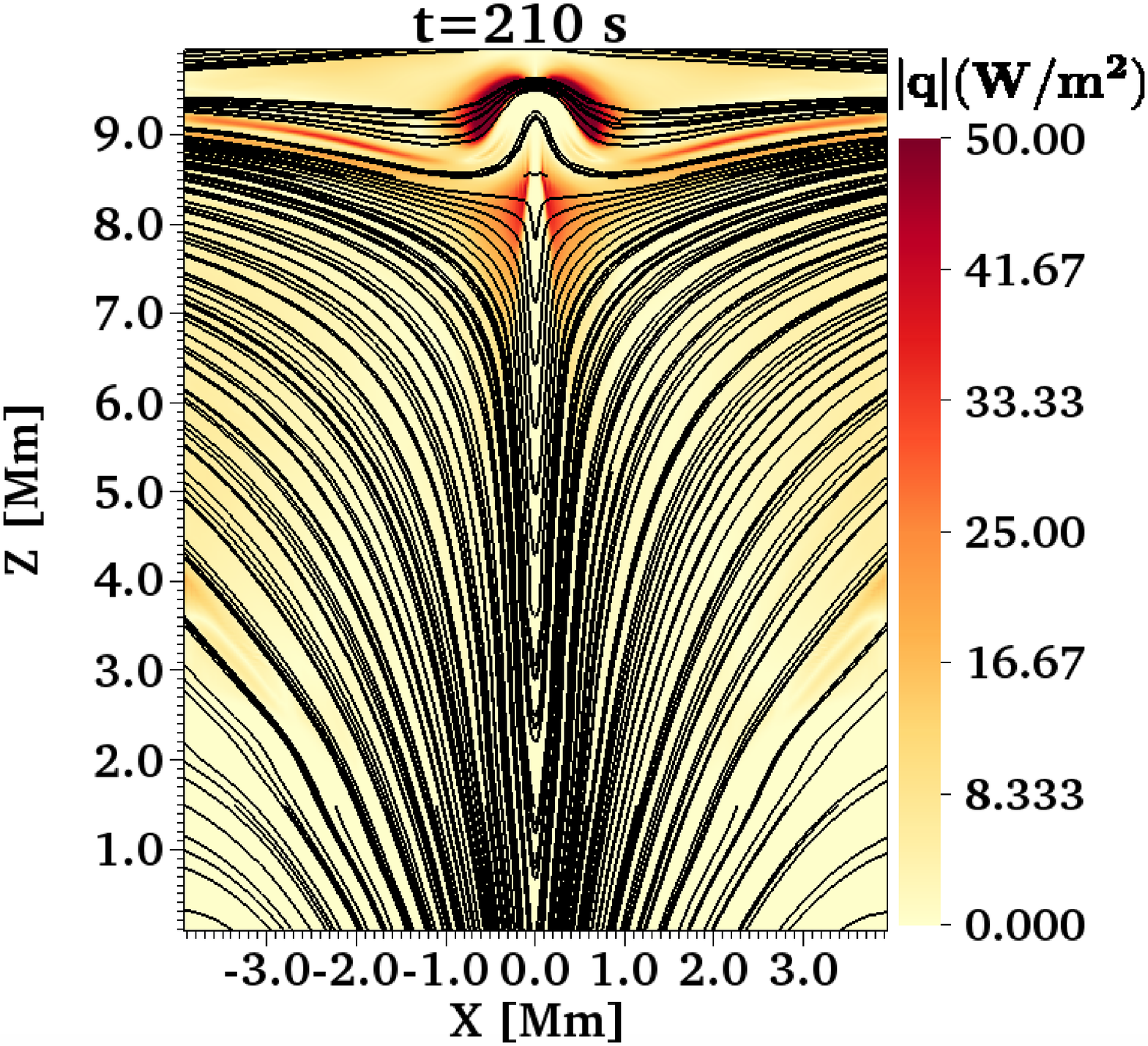}\\
\centerline{\Large \bf   
      \hspace{0.16 \textwidth}  \color{black}{\normalsize{(g)}}
      \hspace{0.2\textwidth}  \color{black}{\normalsize{(h)}}
      \hspace{0.19\textwidth}  \color{black}{\normalsize{(i)}}
         \hfill}
\includegraphics[width=4.3cm,height=5.5cm]{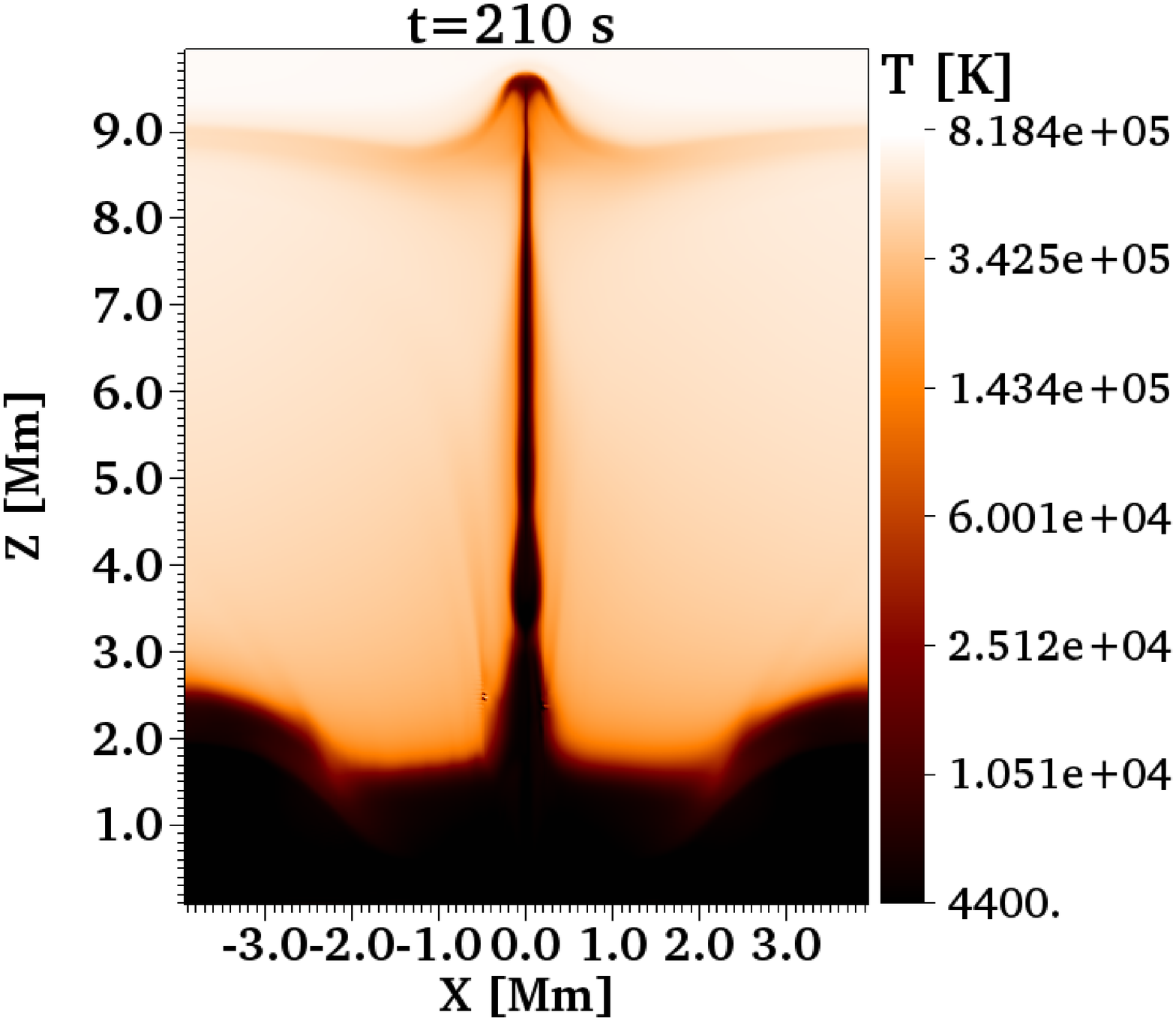}
\includegraphics[width=4.3cm,height=5.5cm]{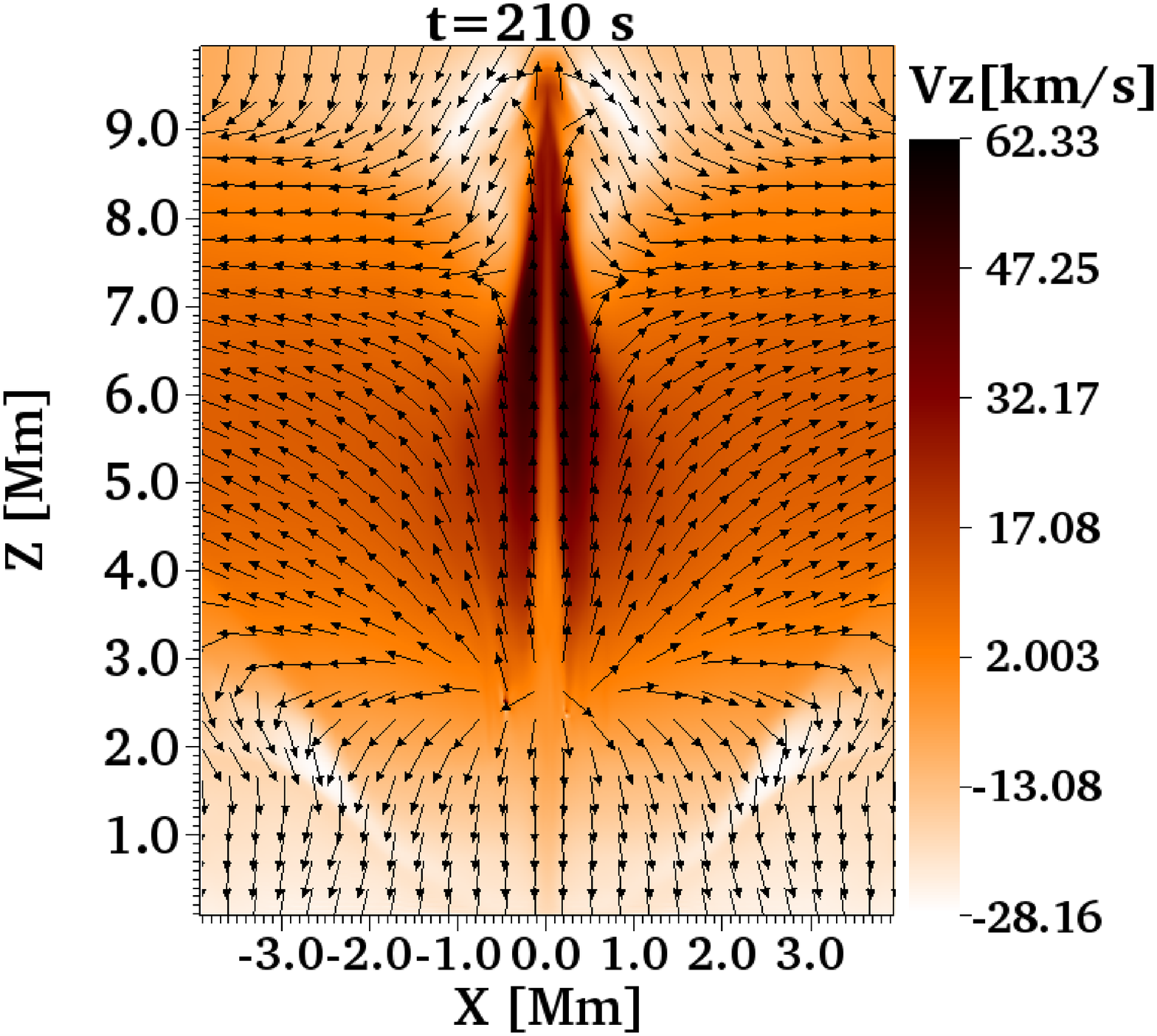}
\includegraphics[width=4.3cm,height=5.5cm]{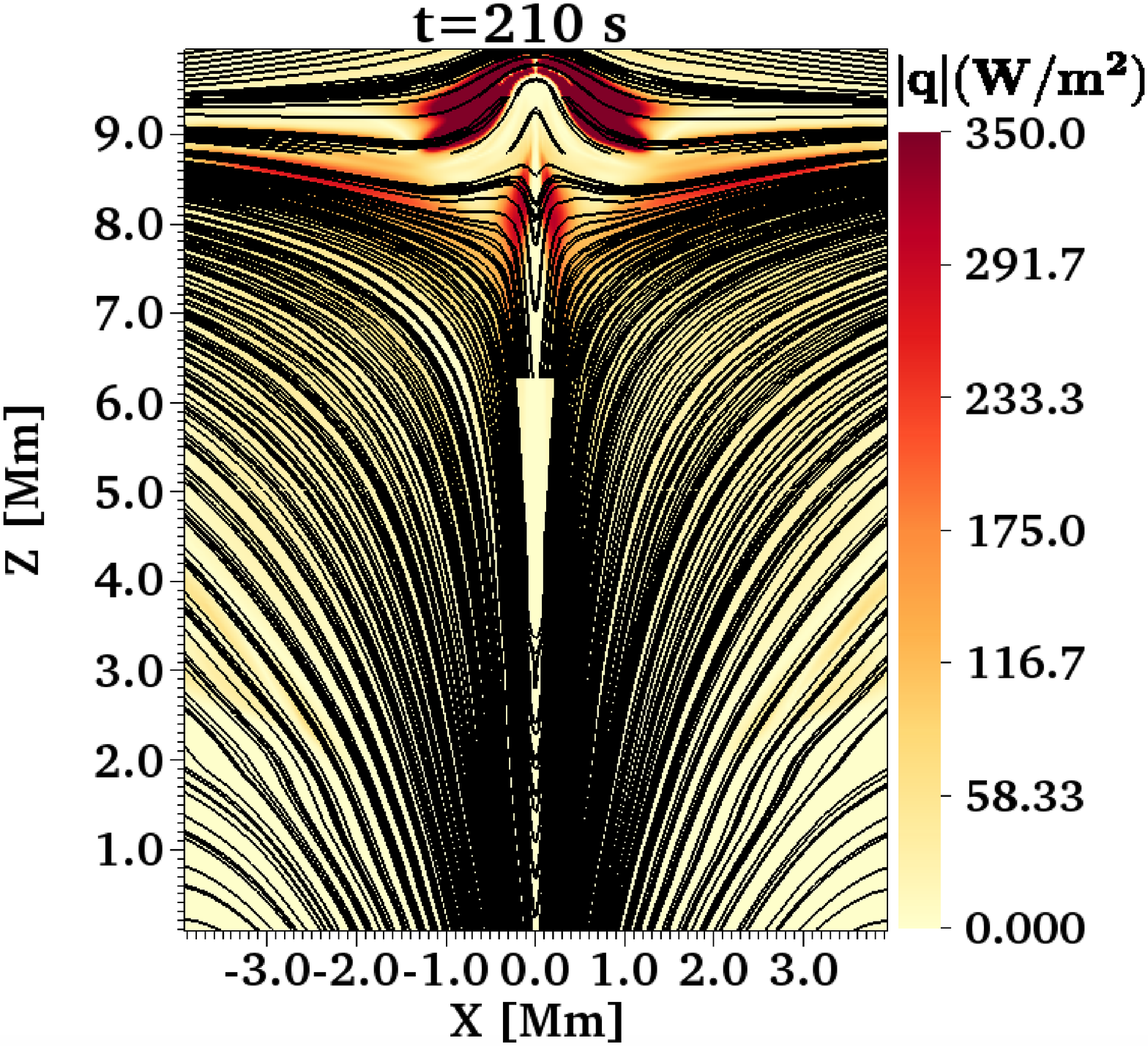}
\caption{From left to right we show snapshots of (i) temperature in Kelvin, (ii) the vertical component of velocity ($v_{z}$ km s$^{-1}$), where the arrows show the velocity field and (iii) the magnitude of the heat flux  $|\bf{q}|$ in W m$^{-2}$ with magnetic field lines. In panels a, b and c we show the results for Run \#1 ($\eta=3.97\times10^{-8}$ and $\kappa=0$) at time $t=210$ s. In the panels d, e and f we show the results for the Run \#6 ($\eta=3.97\times10^{-8}$ and $\kappa=3275$) at time $t=210$ s. Finally, in the panels g, h and i, we show the results for Run \#10 ($\eta=8\times10^{-8}$ and $\kappa=32755$) at time $t=210$ s. Note that as the value of thermal conductivity increases the jet gets thinner closer to the observations. In this figure we only show temperature maps, but the mass density has the same morphology. Heat flux is particularly consistent, notice it is nearly zero at the body of the jet where $\nabla T$ is perpendicular to the field lines, and becomes important at the top of the jet, where $\nabla T$ is more parallel to the shape of the jet-apex and field lines nearly horizontal. It is also seen that heat transfer produces a stream of cold material from the jet-apex to the sides. We use different density of magnetic field lines in each case, to capture some of the differences, specially at the jet-apex.}
\label{fig:2}
\end{figure*}

Of all the runs for simulations and summarized in Table \ref{table1}, we select 3 illustrative ones, e.g. Run \#1: $\eta=3.97\times10^{-8}$, $\kappa=0$, Run \#6: $\eta=3.97\times10^{-8}$, $\kappa=3275$ and Run \#10: $\eta=8\times10^{-8}$, $\kappa=32755$, which are highlighted in light gray. In Figure \ref{fig:2} we show snapshots of the temperature, vertical velocity with the vector field distribution and the magnitude of the heat flux  $|\bf{q}|$ given by equation (\ref{q_along_field_lines}) with magnetic field lines at time $t=210$ s for the three illustrative Runs. For example, in Figures \ref{fig:2}(a), (b) and (c) we show the results for the values $\eta=3.97\times10^{-8}$ and $\kappa=0$, which are practically the same snapshots corresponding to the top of Figure 3 of the paper \cite{2.5Dspicules}, these snapshots will be useful for comparison with cases where thermal conductivity is included. Figure \ref{fig:2}(c) shows the magnetic field lines with $|\bf{q}|=0$, corresponding to $\kappa=0$, for completeness. 

In Figures \ref{fig:2}(d), (e) and (f) we show the results for Run \#6, with the combination of parameters $\eta=3.97\times10^{-8}$ and $\kappa=3275$. In this case according to the temperature map, we can see that the jet is wider in its lower part and thinner in the upper part just below its apex compared to the Run \#1, and reaches a height of 7.5 Mm, which represents 0.2 Mm larger than the jet of Run \#1. In addition, the vertical velocity is higher at the sides of the jet compared to that of Run \#1. The magnitude of the heat flux  $|\bf{q}|$ is high at the top of the jet-apex, and at the sides beneath the jet-apex, while inside the jet structure the value tend to zero. The reason is that ${\bf q}$ is a basically a projection of $\nabla T$ along the field lines, then it is maximum when $\nabla T$ is large has an important component parallel to field lines. At the top of jet-apex the gradient of temperature is nearly radial (i.e perpendicular to the ball shape of the jet-apex) as seen in Fig. 2(a) and magnetic field lines have a radial component; another case is that heat flux is nearly zero aside the jet from $z$ between 2 to 7 Mm because the gradient of $T$ is nearly horizontal whereas the magnetic field lines are nearly vertical and therefore effect of thermal conductivity is minimal.   

In Figures \ref{fig:2}(g), (h) and (i) we show the results for Run \#10 with  values $\eta=8\times10^{-8}$ and $\kappa=32755$, which is the combination that includes the highest value of thermal conductivity. In this case we can also see that the bottom of the jet is about 0.1 Mm wider than the bottom of the jet of Run \#1 and jet-apex is 0.28 Mm thinner compared to the Run \#1, and 0.10 Mm thinner compared to the Run \#6. This jet reaches a height of about 7.7 Mm, which is 0.4 Mm higher compared to Run \#1 and 0.2 Mm higher compared to Run \#6, and in the same way it is seen that the vertical speed is greater aside the jet. In Figure \ref{fig:2}(i), $|\bf{q}|$ is higher at the at the top of the jet, similar to Figure \ref{fig:2}(f), however in this case $|\bf{q}|$ reaches higher values, which correspond with  bigger $\kappa$ used in this numerical run. The jet during development reaches maximum speeds of the order 100 km s$^{-1}$ at early times, between 0 and 50 s, while at later times after 60s the speed decreases to values up to 15-30 km s$^{-1}$. Therefore in the first stage of the jet's evolution the speeds are comparable to those of Type II spicules, whereas at later times the speeds are smaller than the lower limit of the observed velocities. Similarly, the vertical speed of our jets have similarities with the observed velocities of the Rapid Redshifted and Blueshifted Excursions (RREs, RBEs), which are in the range of 50-150 km s$^{-1}$ \citep{Langangen_et_al_2008}. In addition, the vector field shows the appearance of vorticity near the top of the jets. According to the results reported in Table \ref{table1}, the width of the jet-apex at the maximum height vary in the range 0.8-1.1 Mm, which is four times greater than the width of 0.25 Mm that has been observed in RREs and RBEs spicule features \citep{Kuridze_et_al_2015}, however the widths of the RREs and RBEs are of the entire observed structure. In fact, if we estimate the width of the jet structure from our simulations, we obtain that the width varies in the range 0.2-0.6 Mm from the bottom of the jet to below its apex, these widths are close to the observed values. Regarding  the spicules observed at the limb, it has been estimated cross-sectional widths in the range 0.27-0.36 Mm \citep{Sharma_et_al_2018}, which are again smaller than the width of the jet-apex, but they are close to the widths estimated in the entire jet structure of our simulations.  Similarly to the results obtained in \cite{2.5Dspicules}, in this paper we find that jets show a special feature at the apex with a bulb possibly related to the formation of a Kelvin-Helmholtz (KH) type of instability. However as it was shown in \cite{2.5Dspicules}, this instability is suppressed by the magnetic field. 

\begin{figure*}
\centering
\centerline{\Large \bf   
      \hspace{0.14 \textwidth}  \color{black}{\normalsize{(a)}}
      \hspace{0.27 \textwidth}  \color{black}{\normalsize{(b)}}
      \hspace{0.275 \textwidth}  \color{black}{\normalsize{(c)}}
         \hfill}
\includegraphics[width=5.8cm,height=4.2cm]{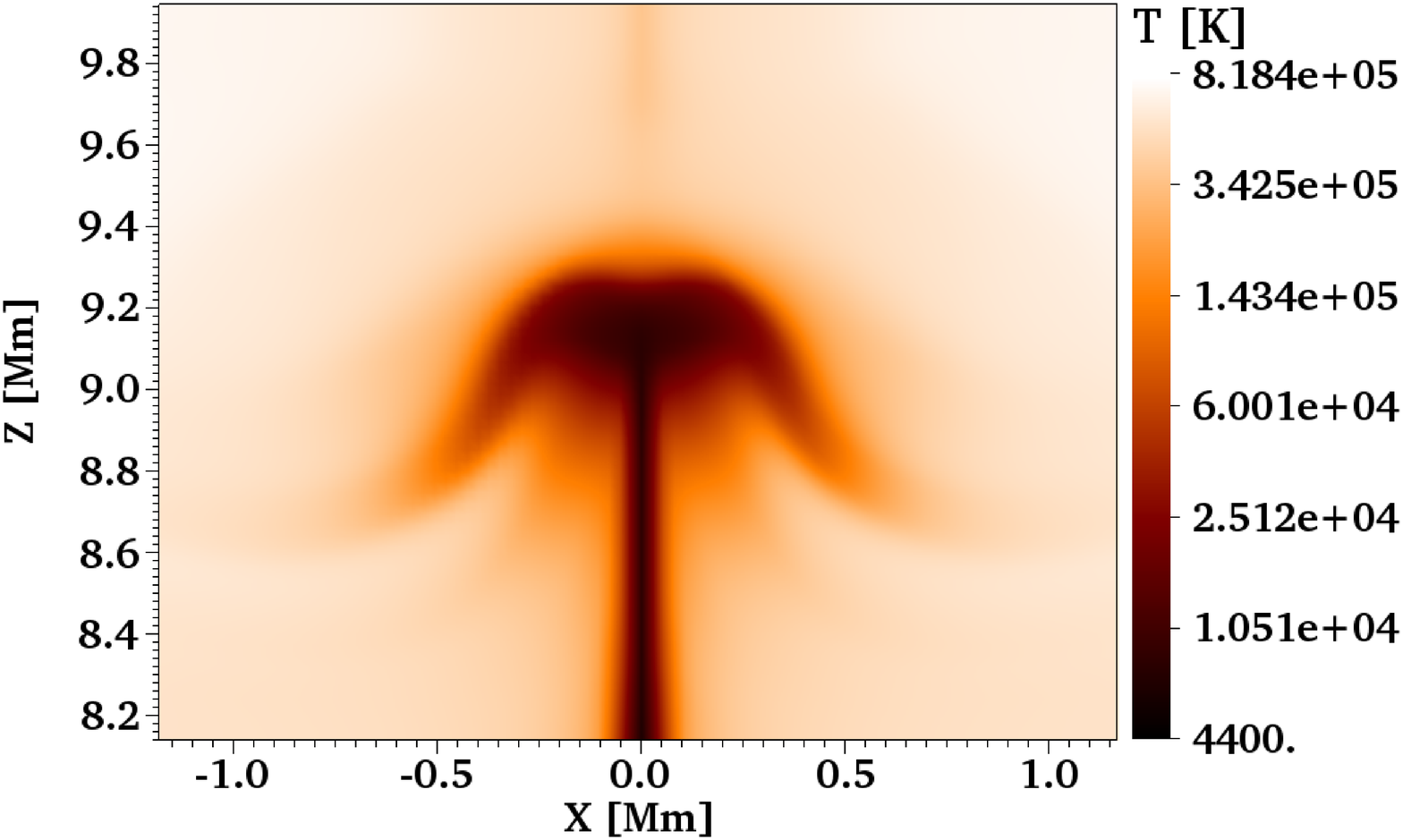}
\includegraphics[width=5.8cm,height=4.2cm]{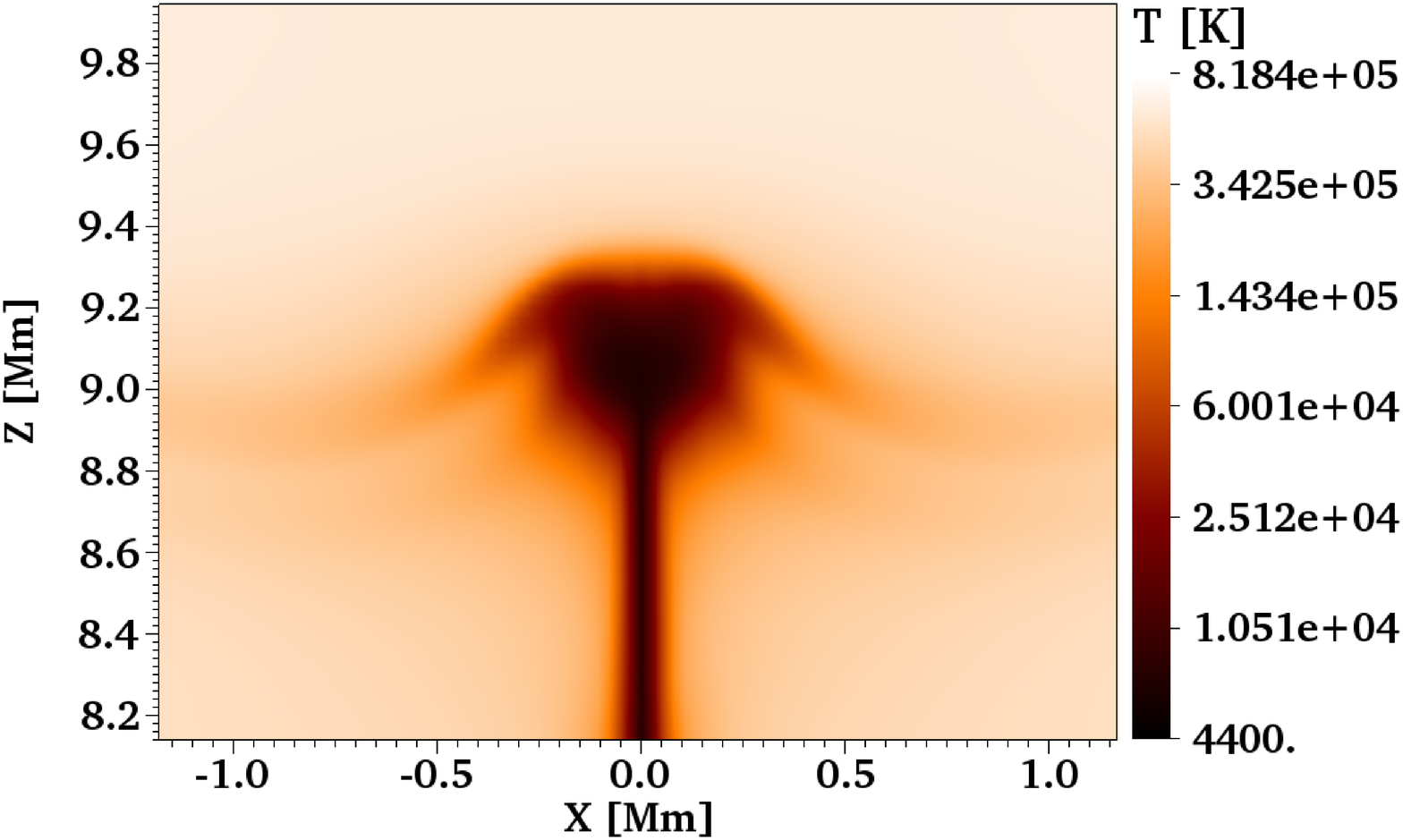}
\includegraphics[width=5.8cm,height=4.2cm]{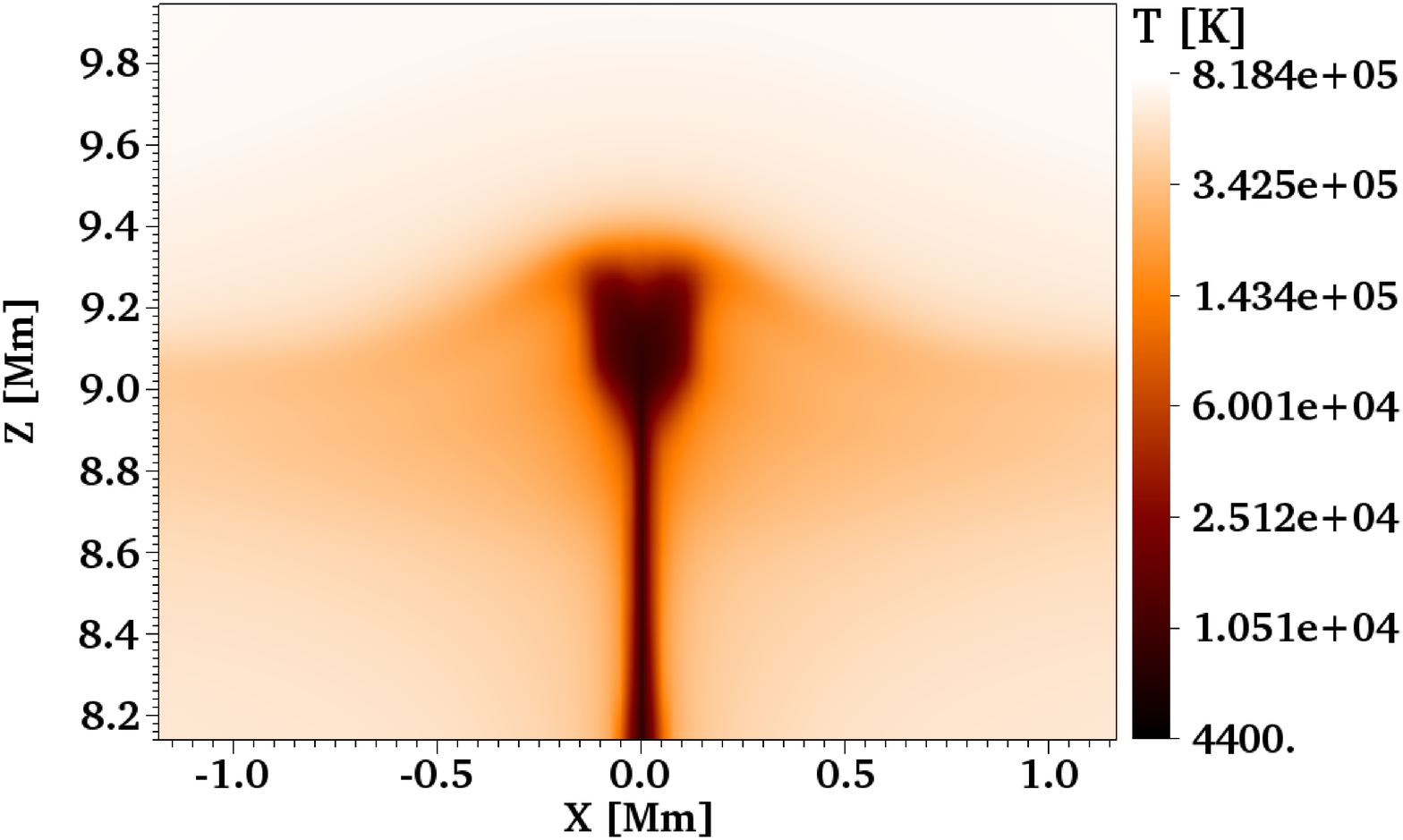}\\
\centerline{\Large \bf   
      \hspace{0.145 \textwidth}  \color{black}{\normalsize{(d)}}
      \hspace{0.275 \textwidth}  \color{black}{\normalsize{(e)}}
      \hspace{0.28 \textwidth}  \color{black}{\normalsize{(f)}}
         \hfill}
\includegraphics[width=5.8cm,height=4.2cm]{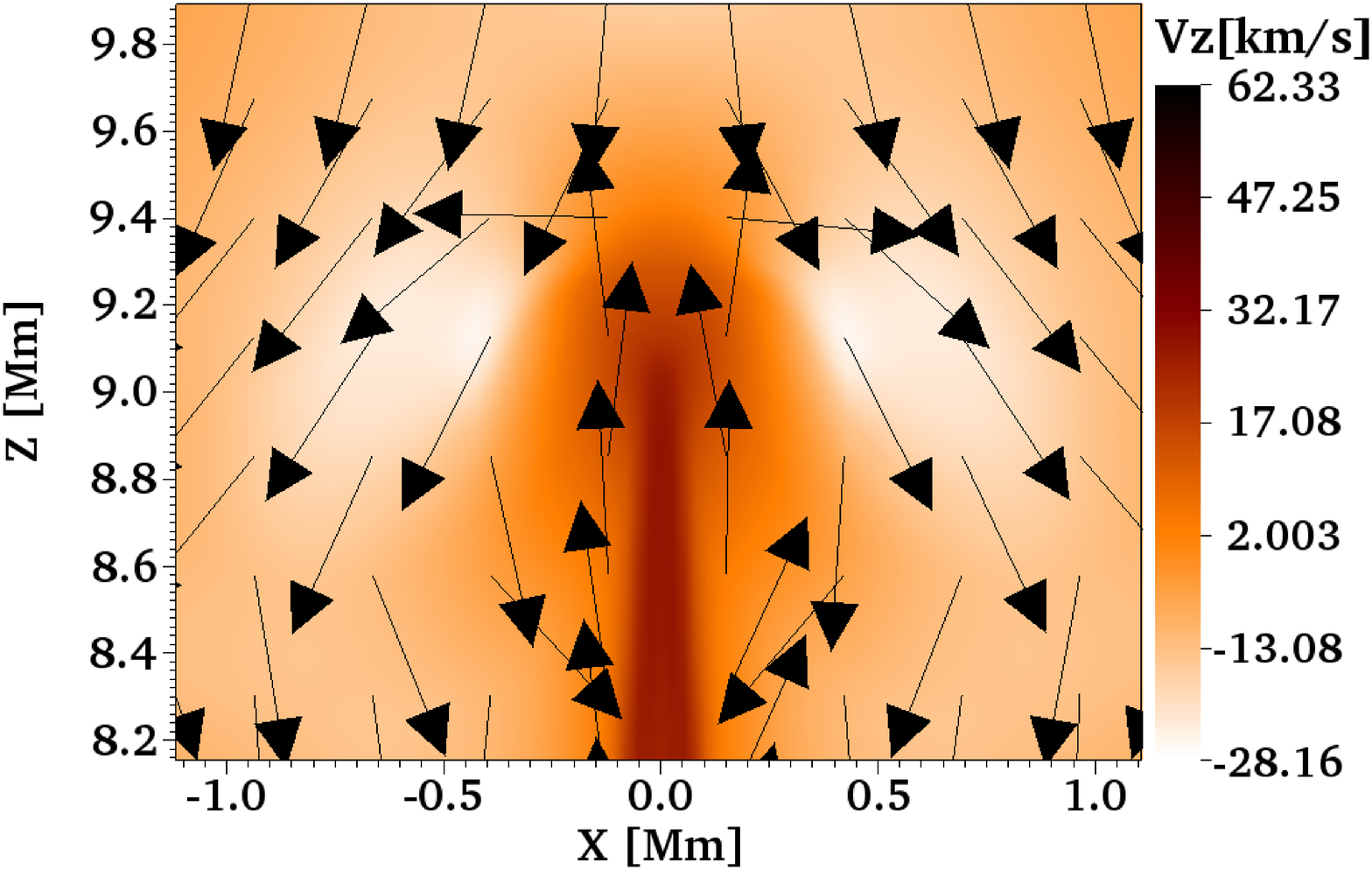}
\includegraphics[width=5.8cm,height=4.2cm]{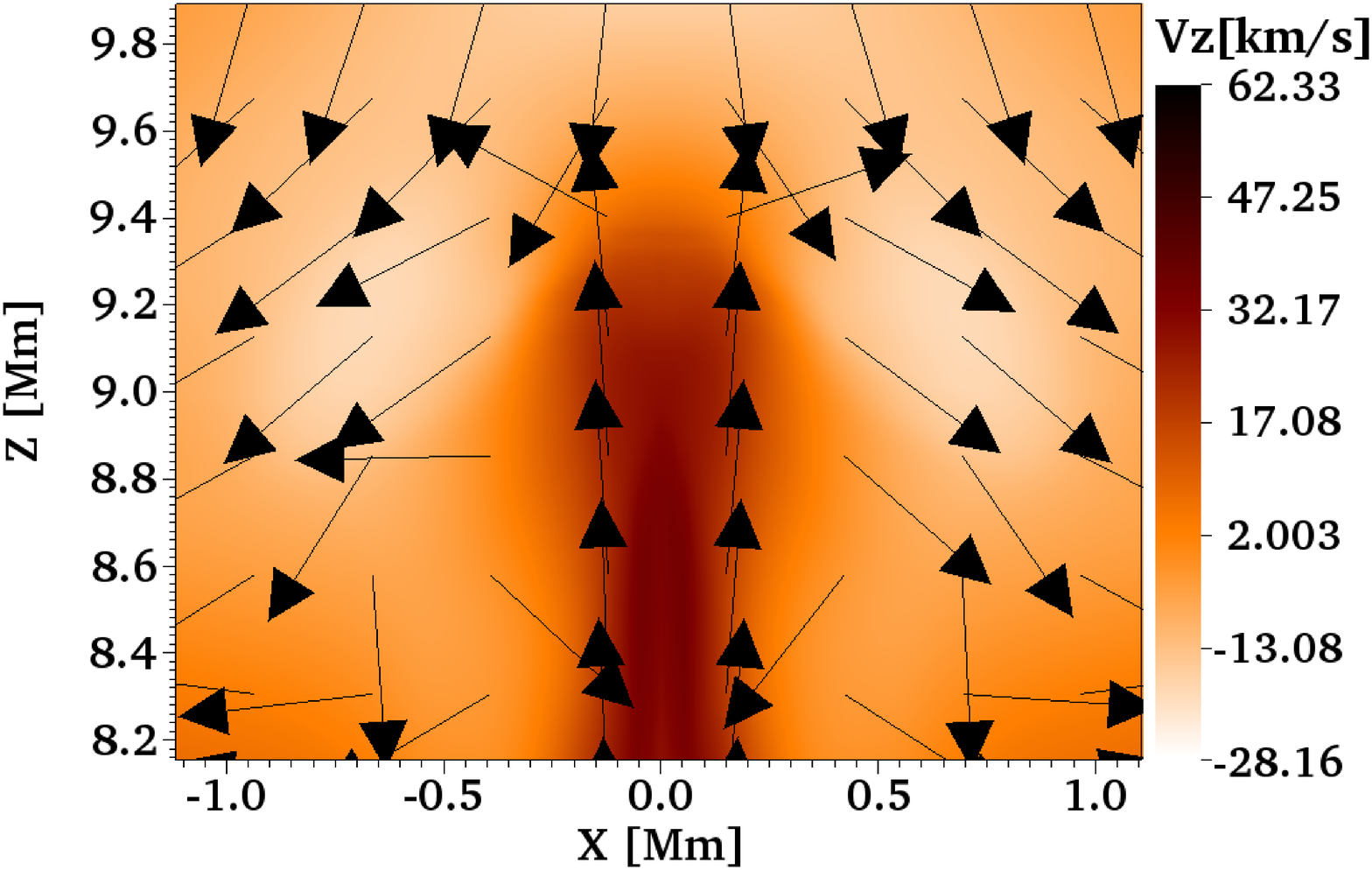}
\includegraphics[width=5.8cm,height=4.2cm]{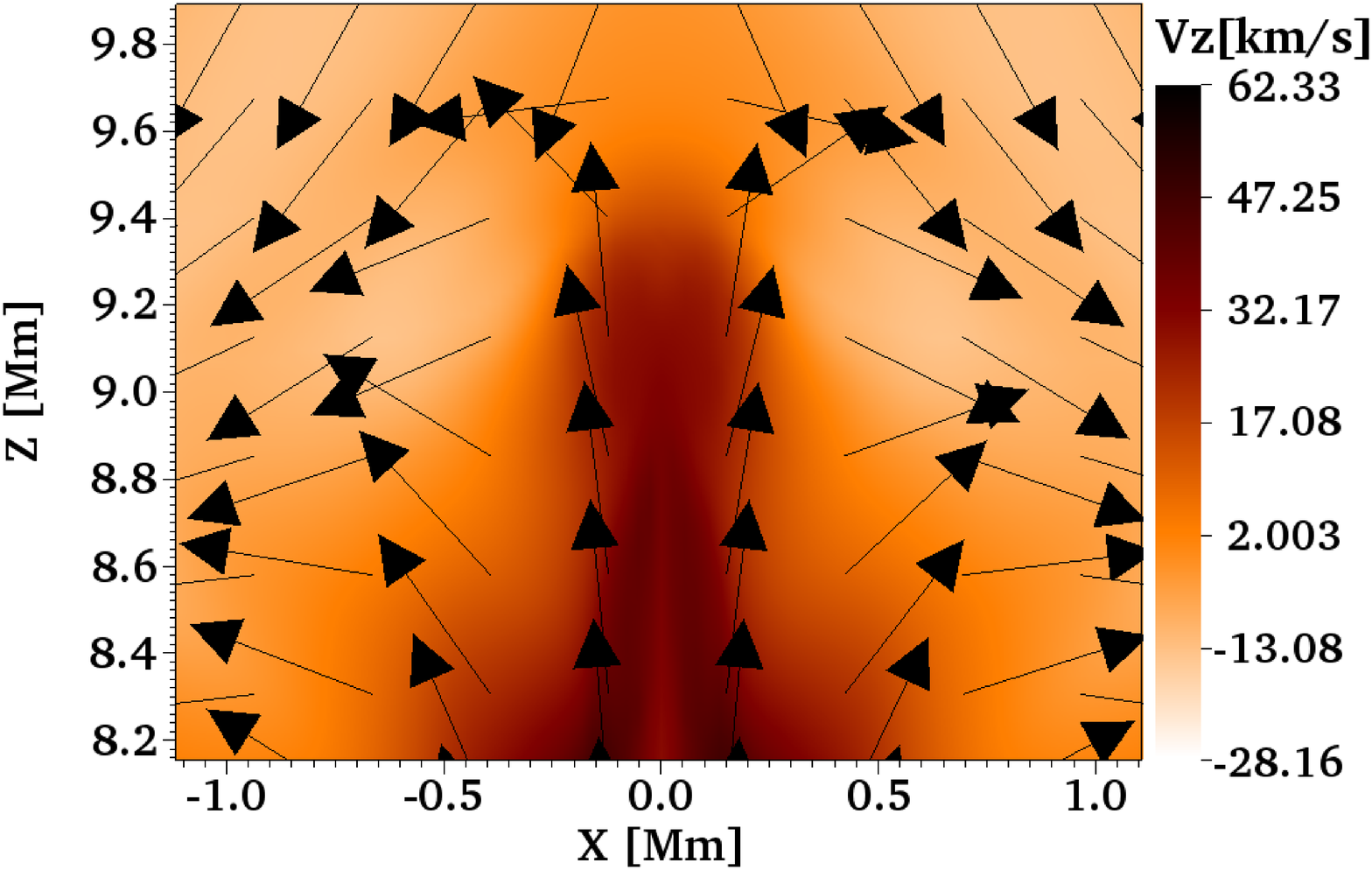}\\
\centerline{\Large \bf   
      \hspace{0.14 \textwidth}  \color{black}{\normalsize{(g)}}
      \hspace{0.275 \textwidth}  \color{black}{\normalsize{(h)}}
      \hspace{0.28 \textwidth}  \color{black}{\normalsize{(i)}}
         \hfill}
\includegraphics[width=5.8cm,height=4.2cm]{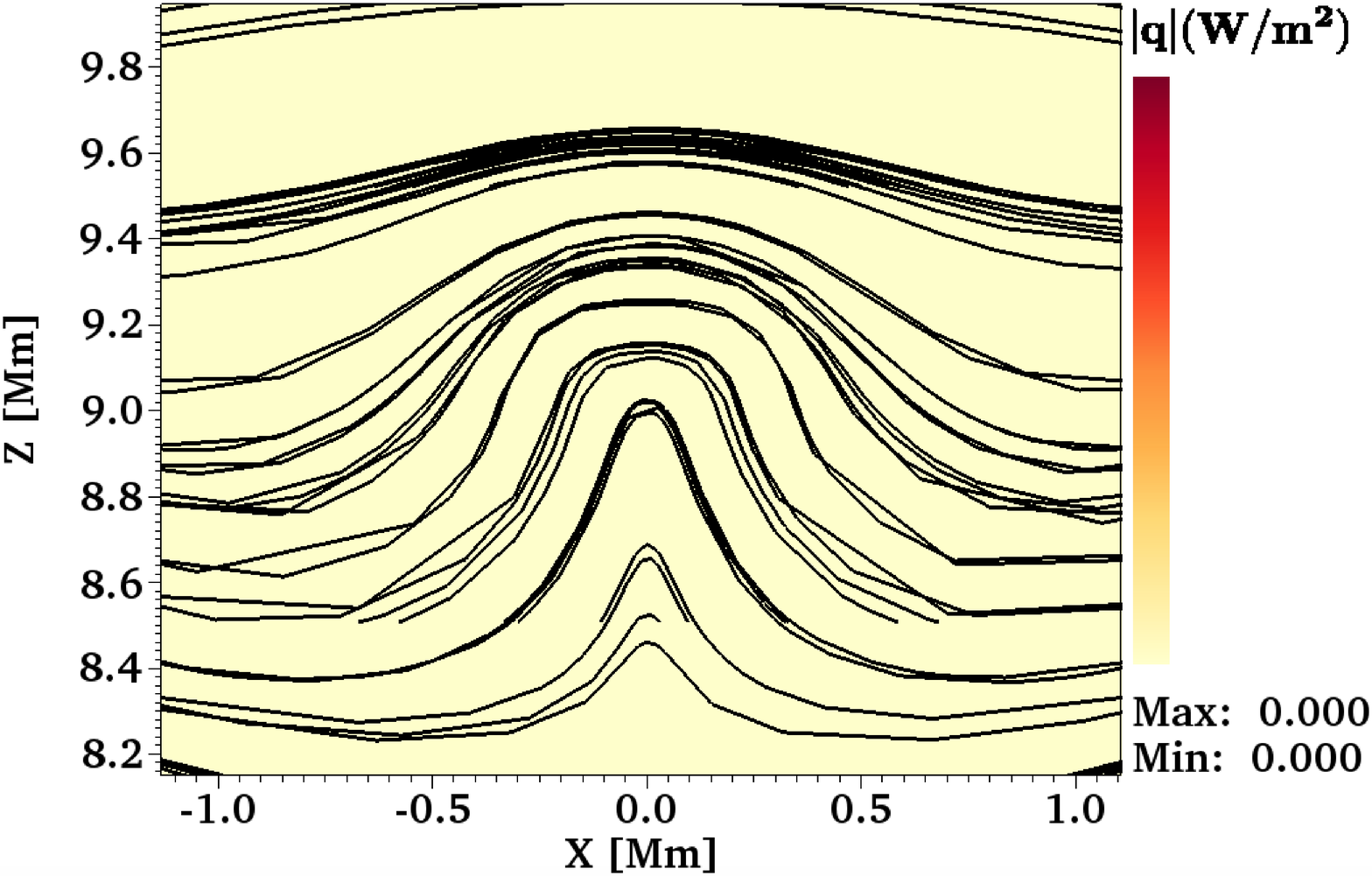}
\includegraphics[width=5.8cm,height=4.2cm]{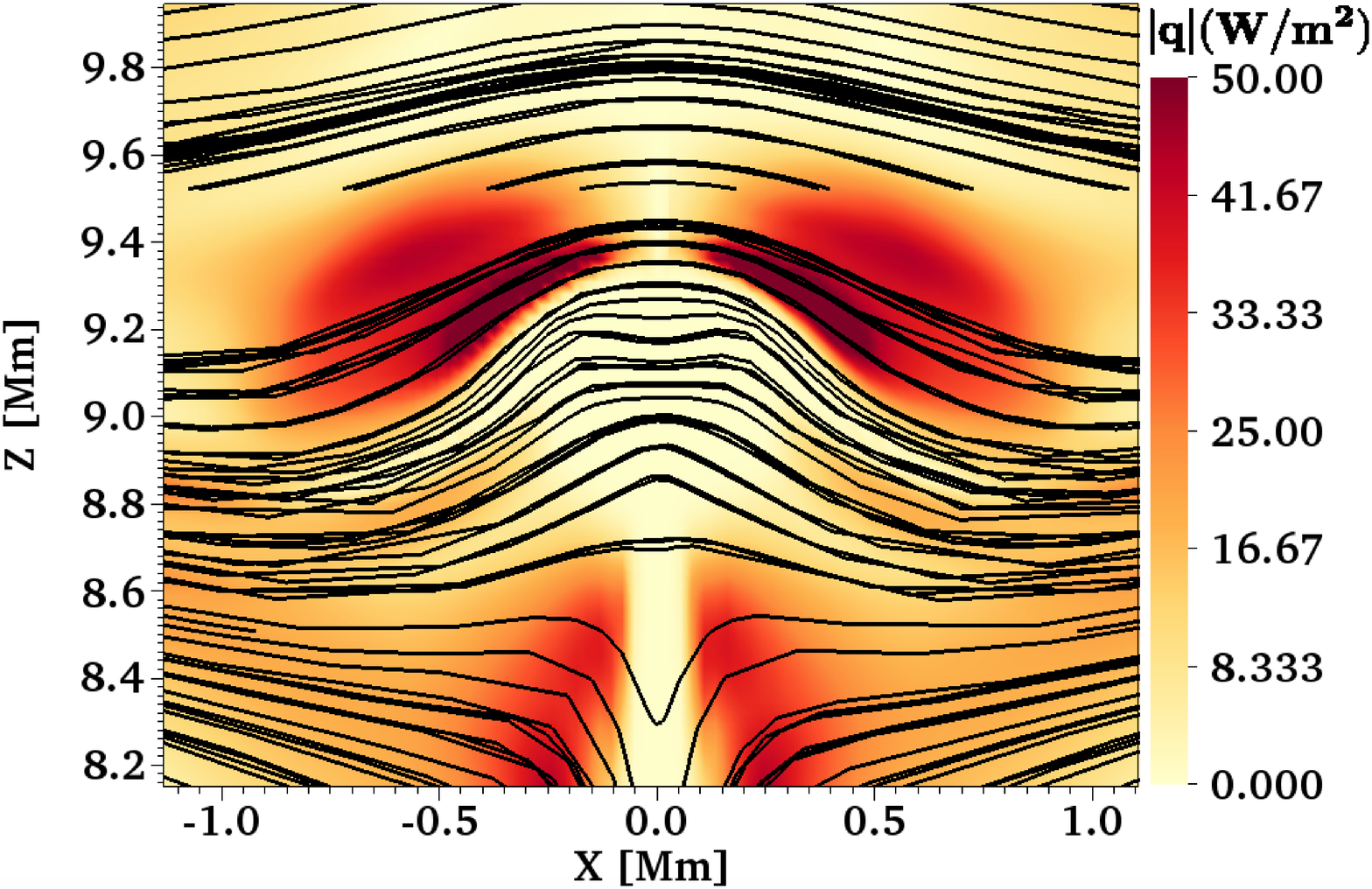}
\includegraphics[width=5.8cm,height=4.2cm]{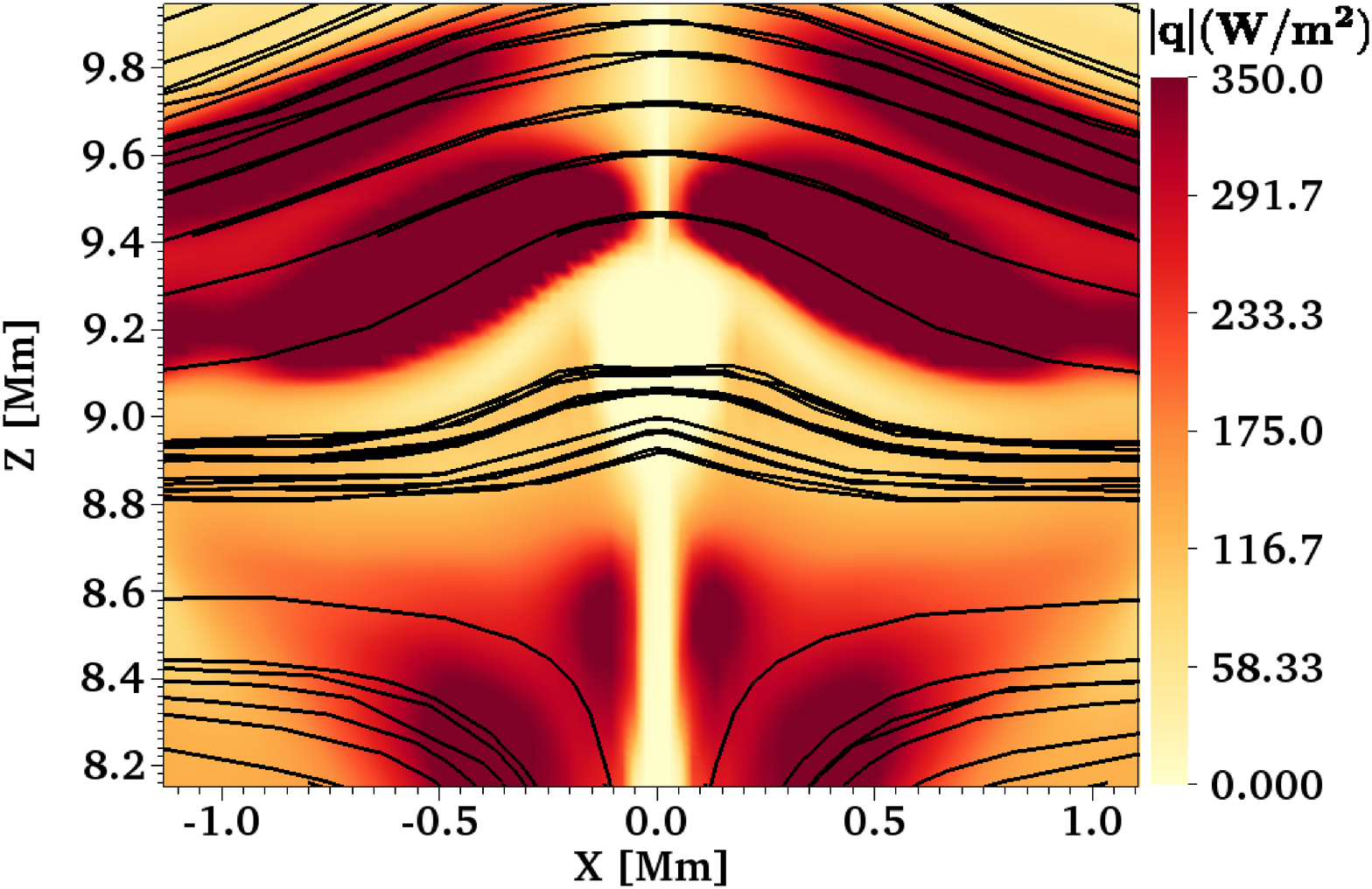}\\
\centerline{\Large \bf   
      \hspace{0.14 \textwidth}  \color{black}{\normalsize{(j)}}
      \hspace{0.28 \textwidth}  \color{black}{\normalsize{(k)}}
      \hspace{0.285 \textwidth}  \color{black}{\normalsize{(l)}}
         \hfill}
\includegraphics[width=5.8cm,height=4.2cm]{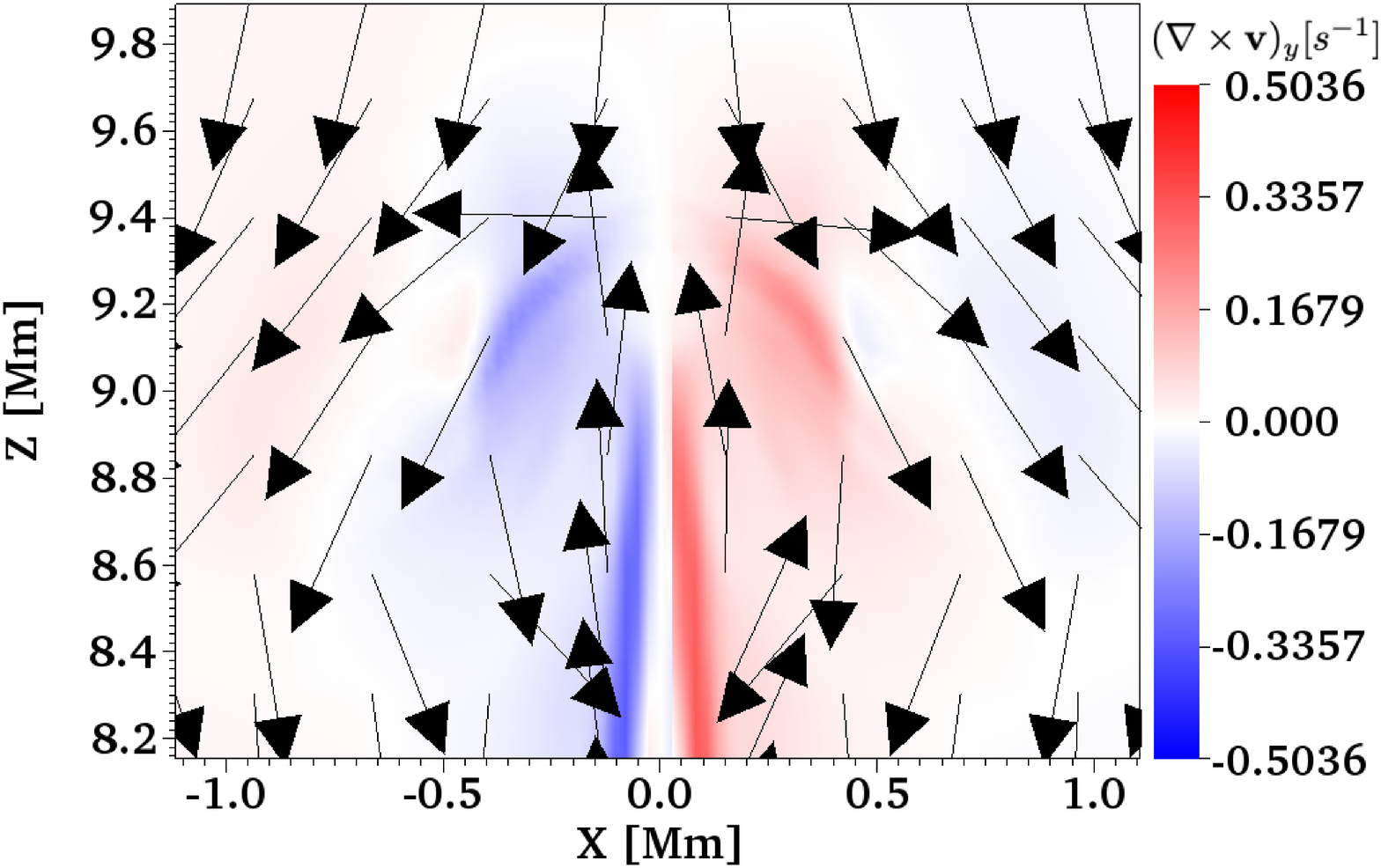}
\includegraphics[width=5.8cm,height=4.2cm]{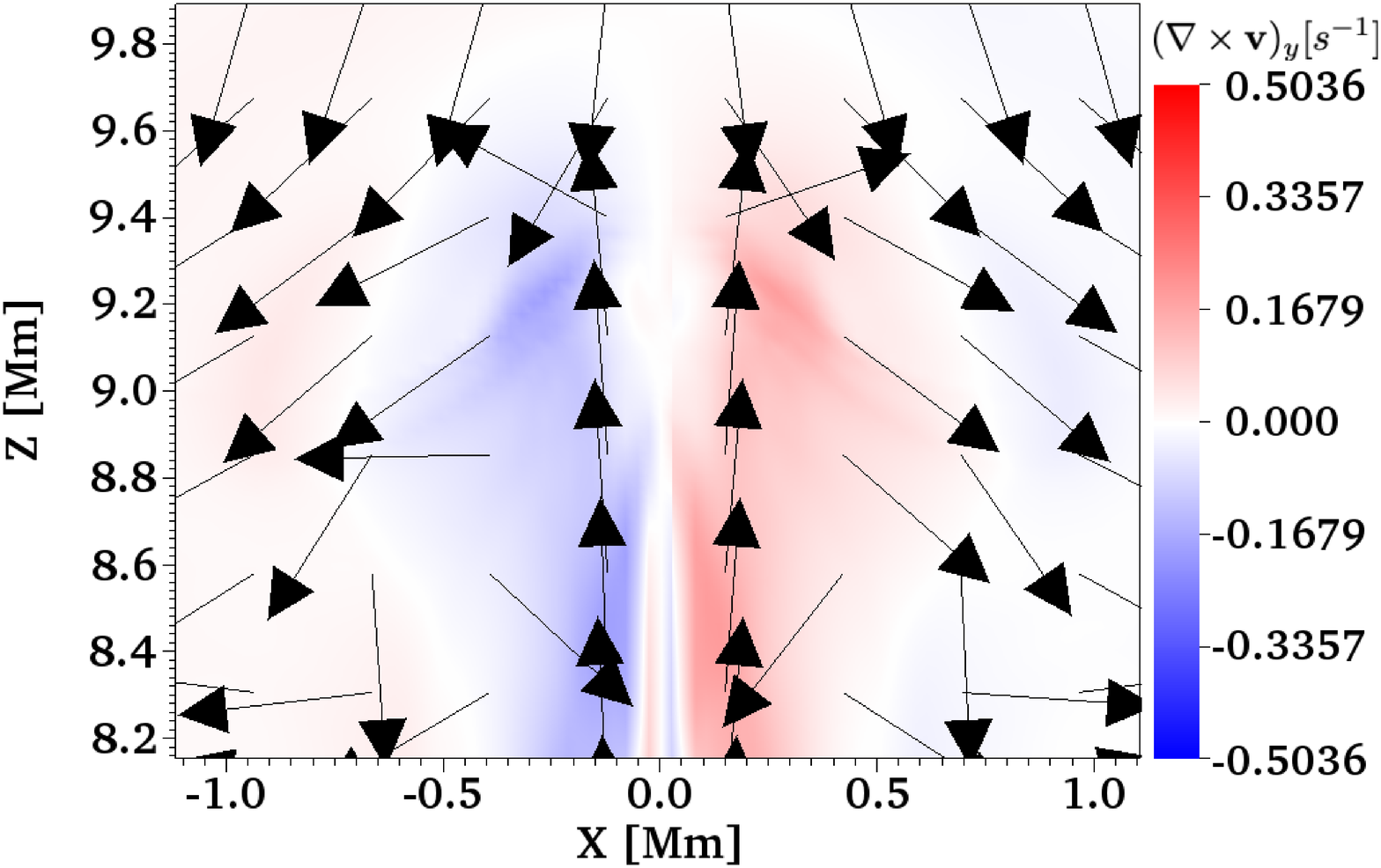}
\includegraphics[width=5.8cm,height=4.2cm]{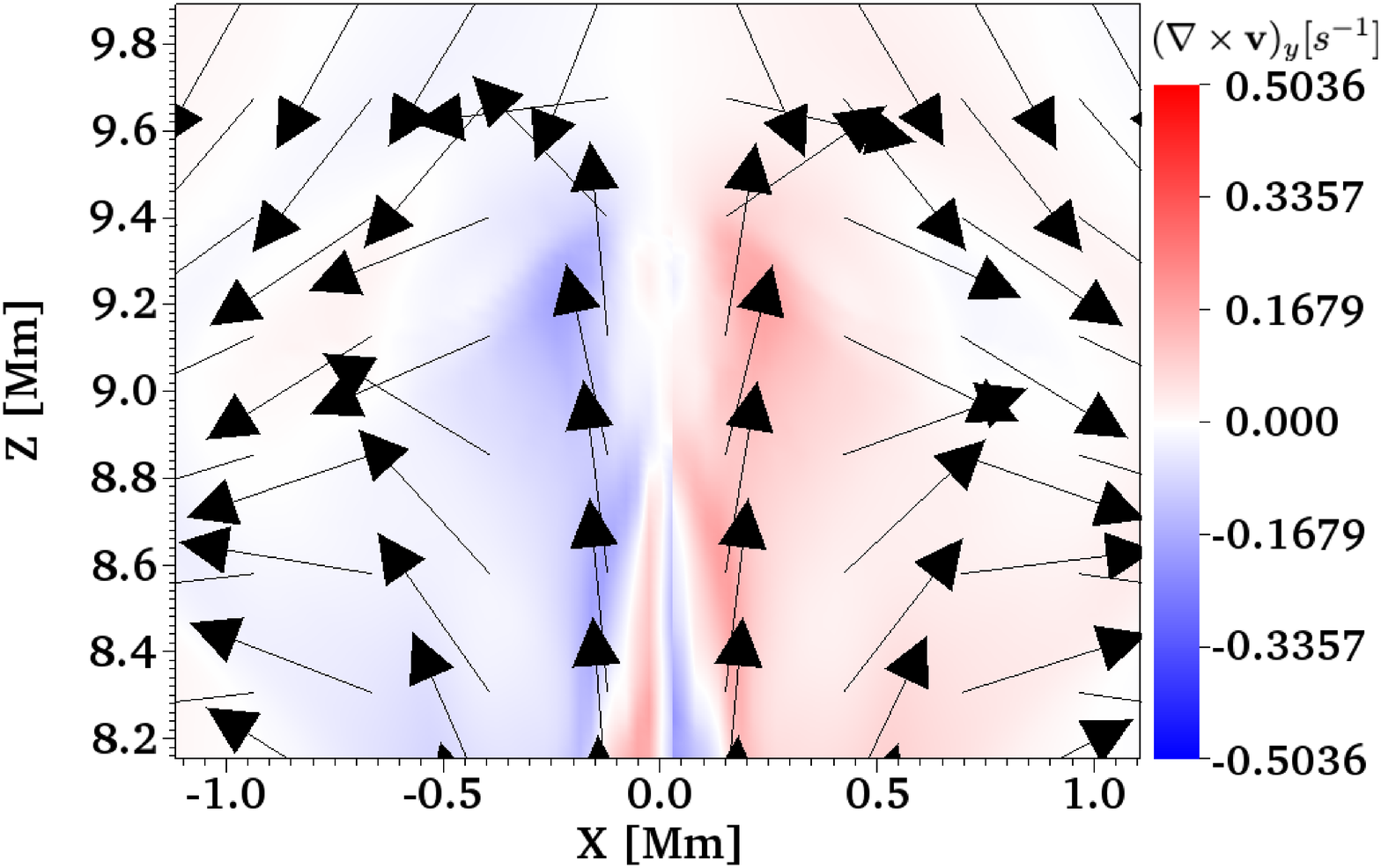}
\caption{From left to right we show a zoom of the region around the jets for Runs \#1, Run \#6 and Run \#10. We show the temperature in Kelvin (panels a, b and c), vertical component of velocity $v_{z}$ in km s$^{-1}$ with the velocity vector field shown as black arrows (panels d, e and f), magnitude of the heat flux  $|\bf{q}|$ in W m$^{-2}$ as well as magnetic field lines (panels g, h and i) and $y$-component of vorticity $(\nabla\times{\bf v})_{y}$ in s$^{-1}$ (panels j, k and l). Note that we only show the temperature, but the mass density has the same morphology. In all the figures the snapshot is taken when the jet is at the same height.}
\label{fig:3}
\end{figure*}

To see more clearly the differences in morphology for the cases shown in Figure \ref{fig:2}, in Figure \ref{fig:3} 
we show a zoom of snapshots of temperature, vertical velocity $v_{z}$ together with the velocity field, magnitude of the heat flux  $|\bf{q}|$ with magnetic field lines and the $y$-component of the vorticity $(\nabla\times{\bf v})_{y}$ with the velocity field at the time when the jets are at the same height. For example, in Figures \ref{fig:3}(a), (b) and (c) we can see that the jets have different morphology when are at the same height, in particular the jet with the highest value of thermal conductivity is smaller and thinner. The cold material develops a horizontal structure connected to the jet-apex that is more notorious for higher $\kappa$, which can also be seen in Figure \ref{fig:2}. In Figures \ref{fig:3}(d), (e) and (f) we show that the vertical component of velocity is higher on the side of the jet for the cases when thermal conductivity is higher. In Figure \ref{fig:3}(g), we show the case of Run \#1 when $|\bf{q}|$ is zero. The magnetic field lines have the shape of the jet shown in Figure \ref{fig:2}(a). In Figures \ref{fig:3}(h) and (i) we show that $|\bf{q}|$ is high near to the top of the jet-apex for Runs \#6 and \#10, where the magnetic field lines have significant horizontal component. We can also see that the magnetic field lines follow the jet structure in Run \#1, whereas for the other two cases the magnetic field lines tend to flatten near the jet-apex. This is consistent with the fact that $\beta<1$ outside the jet and $\beta>1$ inside, where hydrodynamical effects dominate. Finally, in Figures \ref{fig:3}(j), (k) and (l) we can notice that at the sides of the jet-apex, the flux develops vorticity, which apparently does not change with the increase in thermal conductivity. The appearance of the vorticity could be due to the interaction of the cold gas with the hotter plasma in the corona.  

\begin{figure*}
\centering
\centerline{\Large \bf   
      \hspace{0.323 \textwidth}  \color{black}{\normalsize{(a)}}
      \hspace{0.281 \textwidth}  \color{black}{\normalsize{(b)}}
         \hfill}
\includegraphics[width=6.5cm,height=5.5cm]{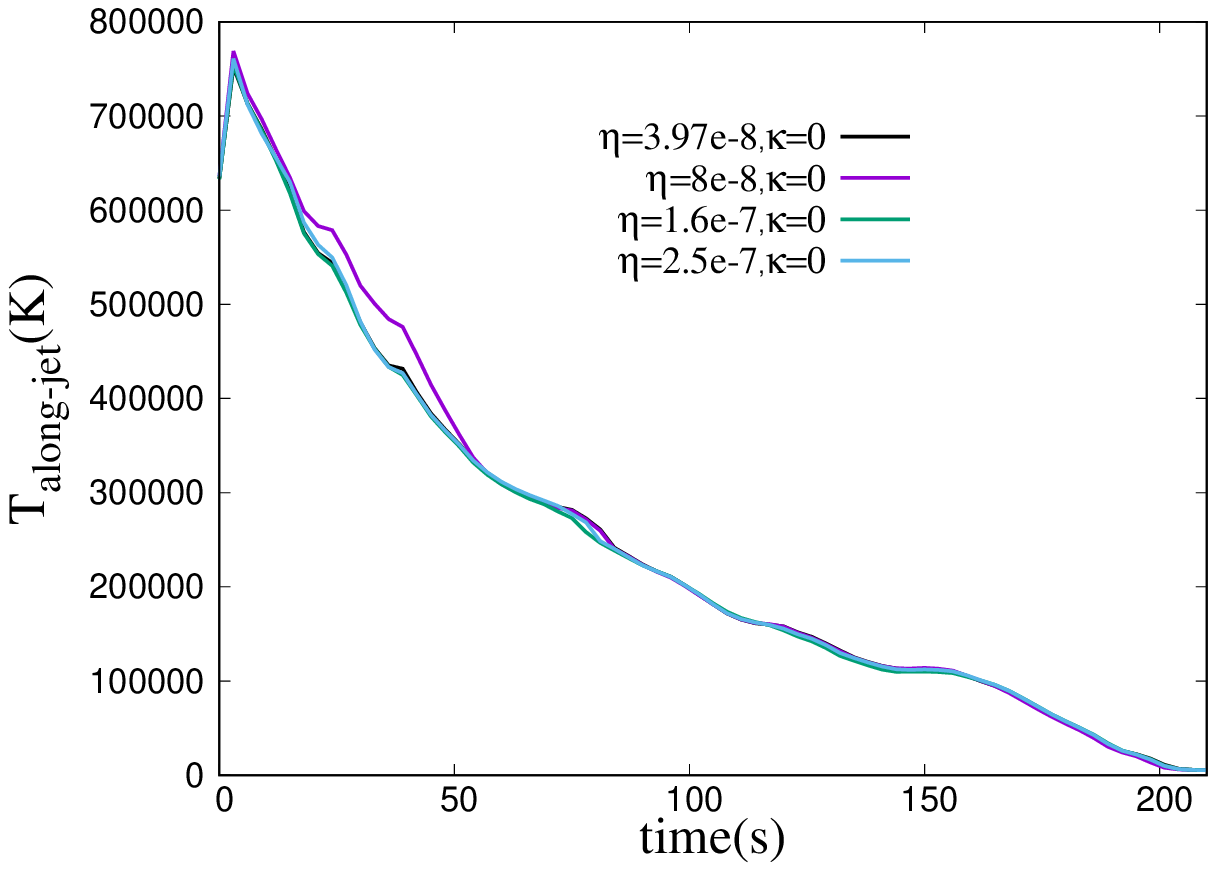}
\includegraphics[width=6.5cm,height=5.5cm]{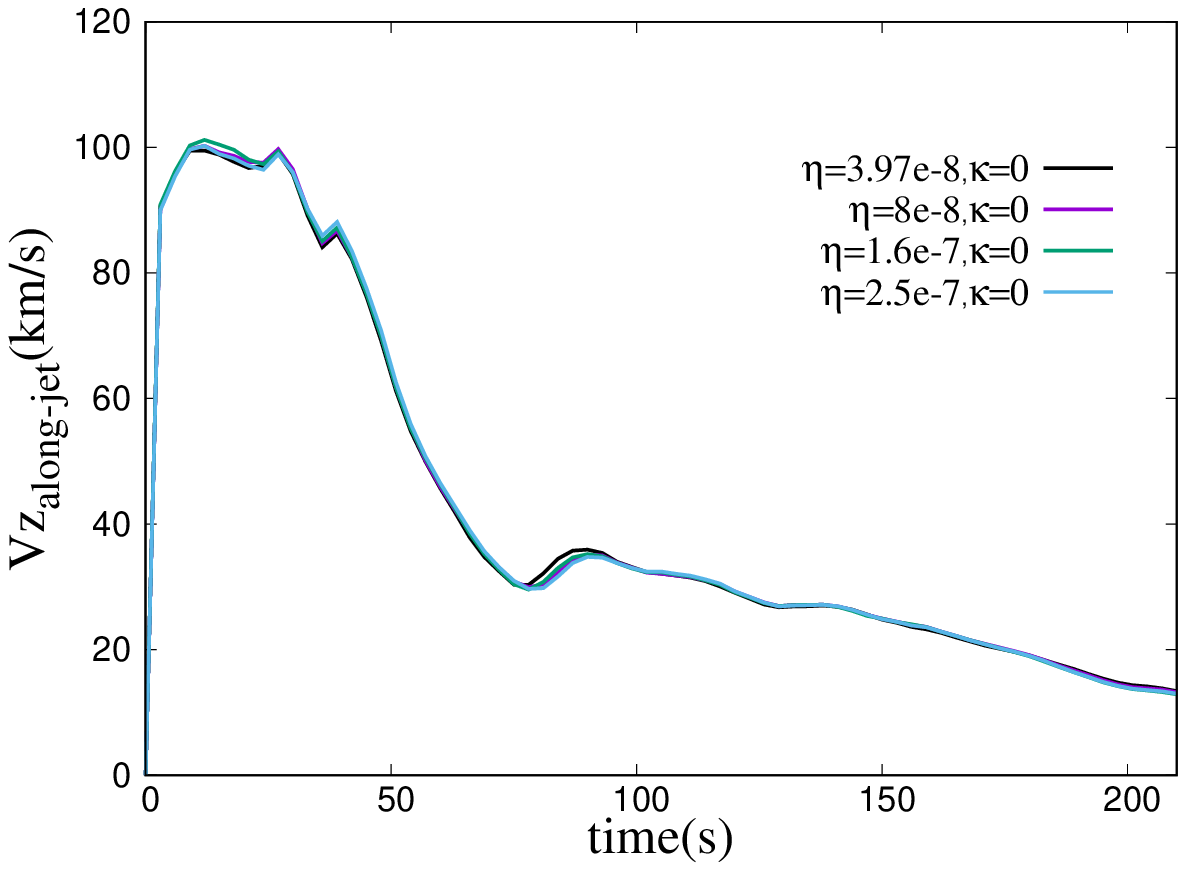}\\
\centerline{\Large \bf   
      \hspace{0.323 \textwidth}  \color{black}{\normalsize{(c)}}
      \hspace{0.281 \textwidth}  \color{black}{\normalsize{(d)}}
         \hfill}
\includegraphics[width=6.5cm,height=5.5cm]{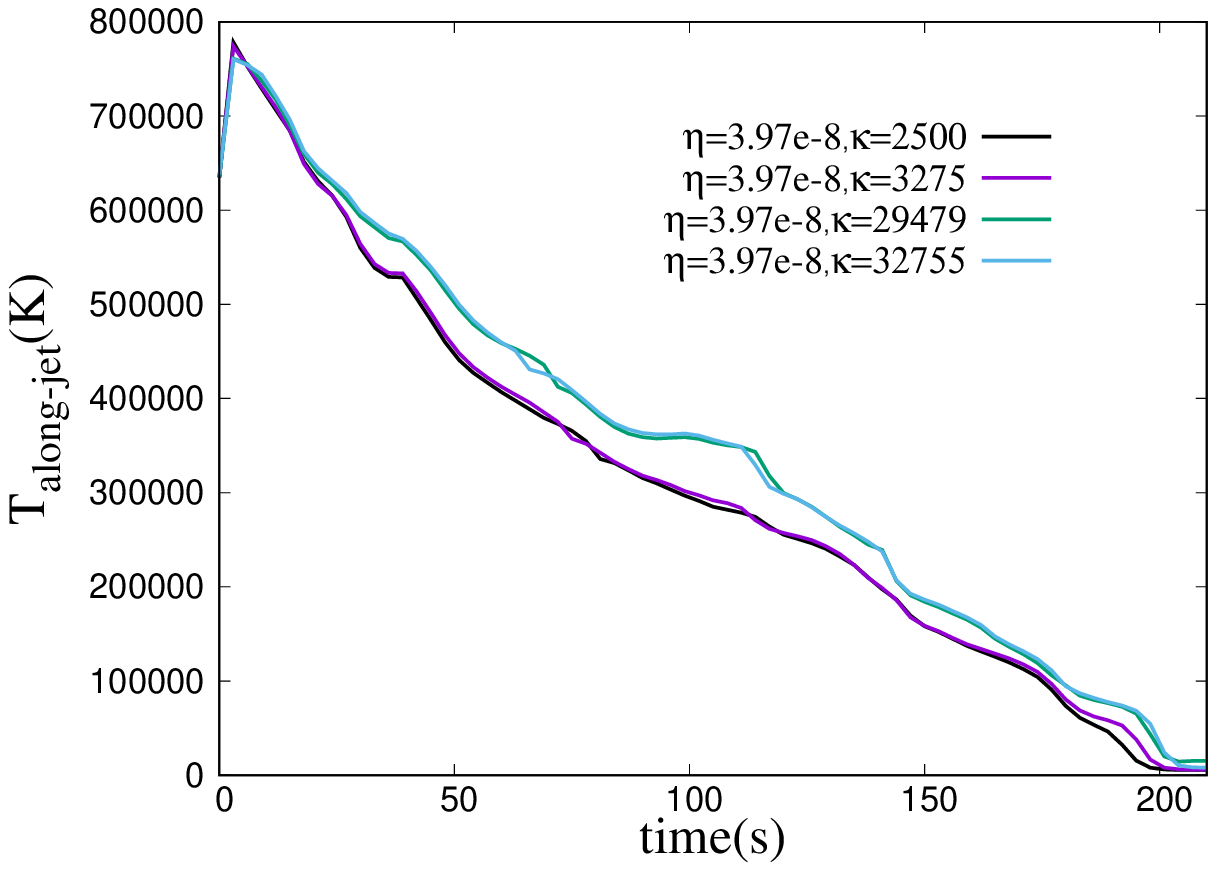}
\includegraphics[width=6.5cm,height=5.5cm]{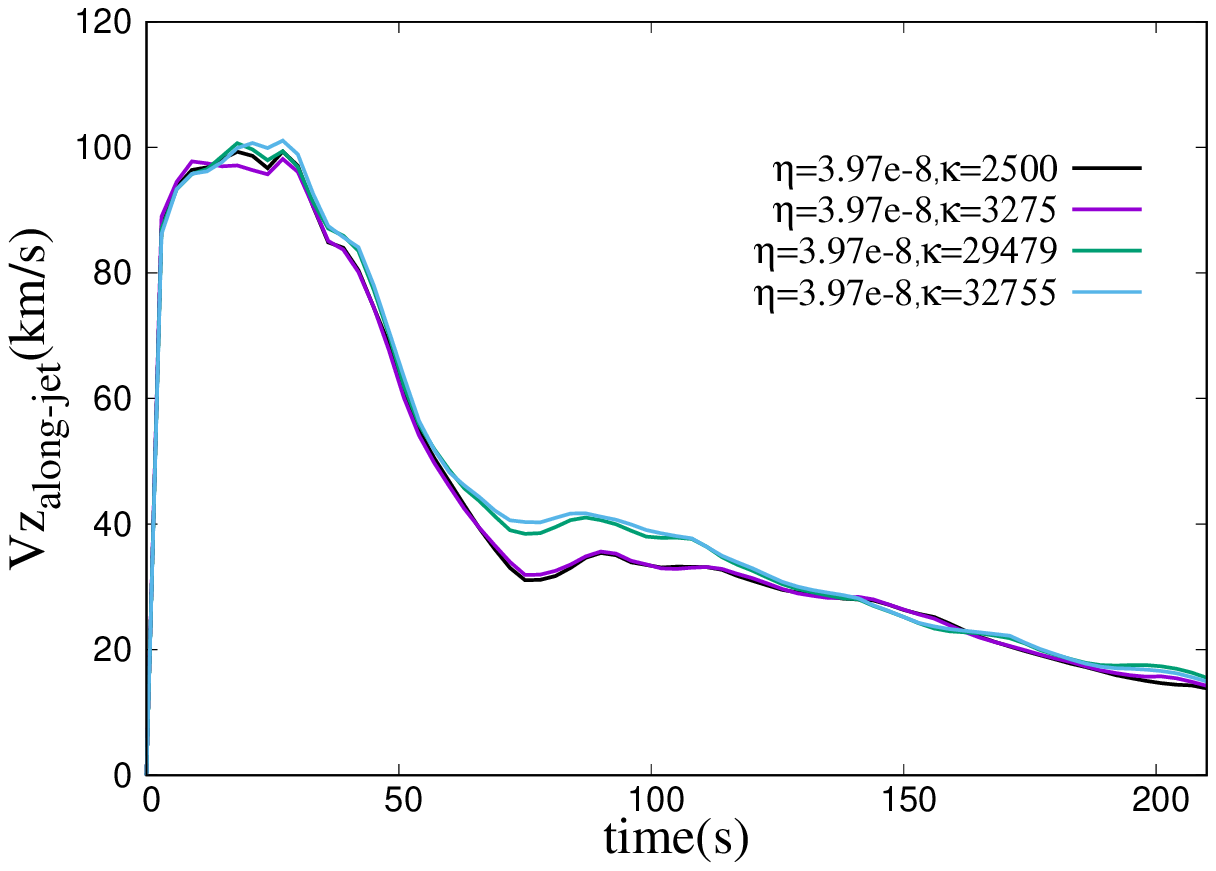}\\
\centerline{\Large \bf   
      \hspace{0.323 \textwidth}  \color{black}{\normalsize{(e)}}
      \hspace{0.29 \textwidth}  \color{black}{\normalsize{(f)}}
         \hfill}
\includegraphics[width=6.5cm,height=5.5cm]{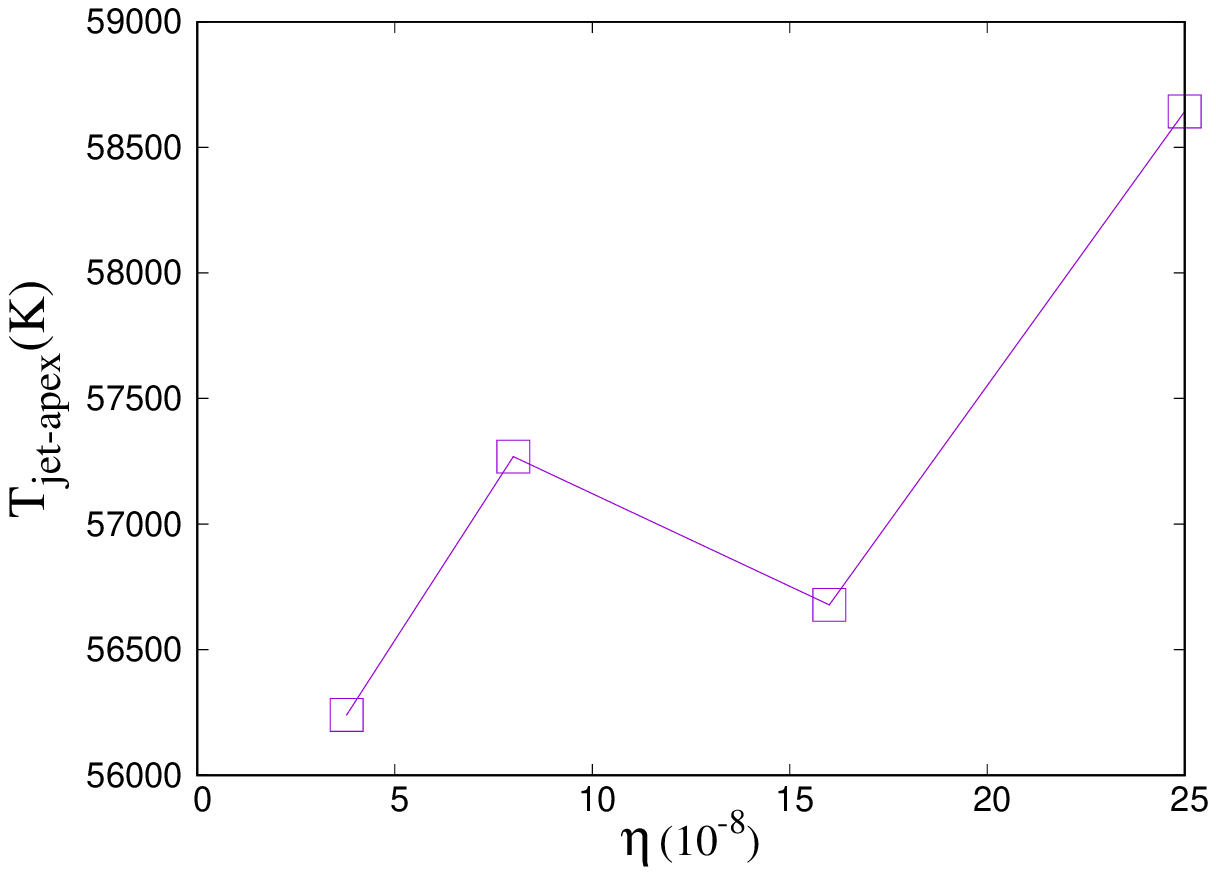}
\includegraphics[width=6.5cm,height=5.5cm]{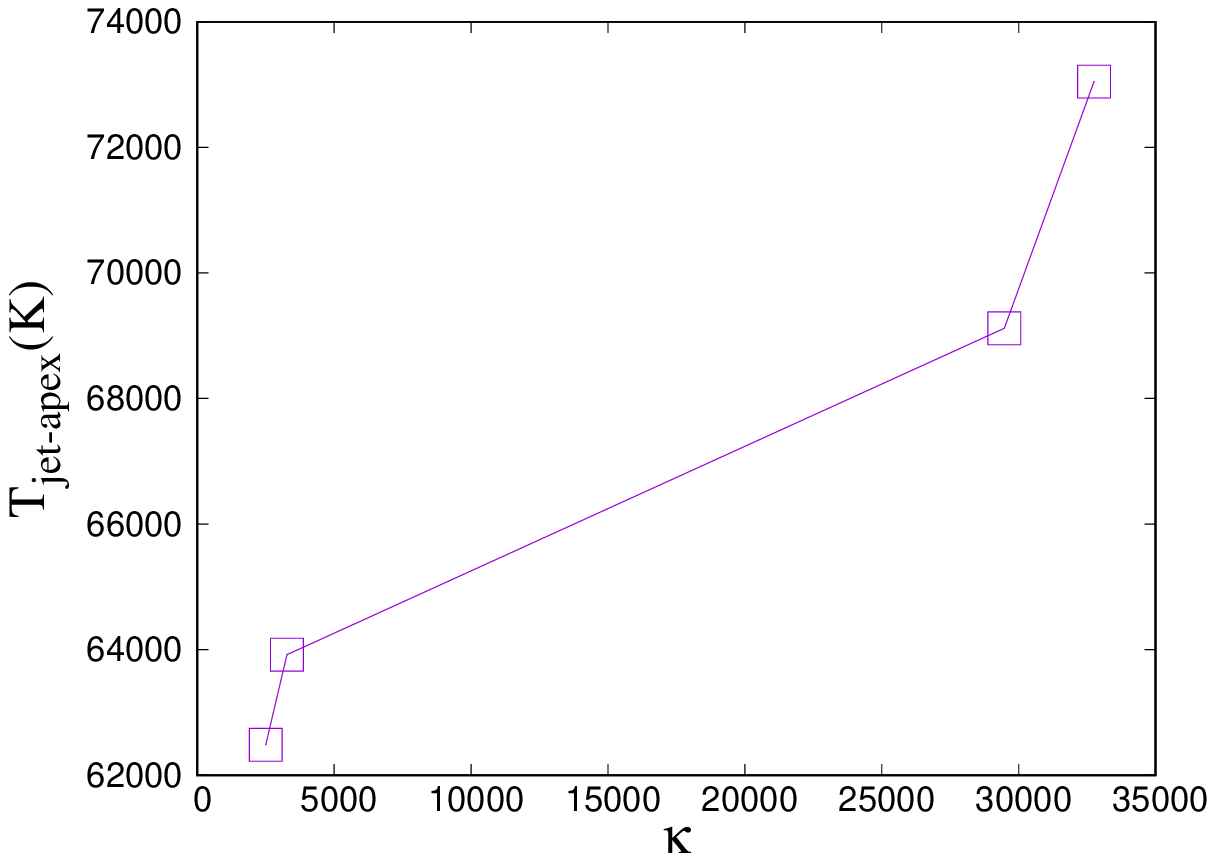}
\caption{Panel (a): temperature along the jet $T_{along-jet}$ in Kelvin as a function of time for different values of $\eta$ and $\kappa=0$. Panel (b): vertical velocity $v_{z}$ along the jet in km s$^{-1}$ as a function of time for different values of $\eta$ and $\kappa=0$. Panel (c): temperature along the jet $T_{along-jet}$ in Kelvin as a function of time for different values of $\kappa$ and $\eta=3.97\times10^{-8}$. Panel (d): vertical velocity $v_{z}$ along the jet in km s$^{-1}$ as a function of time for different values of $\kappa$ and $\eta=3.97\times10^{-8}$. Panel (e): average jet-apex temperature $T_{jet-apex}$ as a function of the $\eta$ at time $t=210$ s. Panel (f): average jet-apex temperature $T_{jet-apex}$ as a function of $\kappa$ at time $t=210$ s.}
\label{fig:4}
\end{figure*}

In order to quantify the influence of resistivity and thermal conductivity, we average the temperature and vertical velocity along the jet and show it as functions of time. We also calculate an average of the jet-apex temperature at time $t=210$ s. For instance, in Figure \ref{fig:4}(a), we show the temperature along the jet as a function of time for different values of $\eta$ and $\kappa=0$. In this case, we notice that temperature does not show significant variations in all cases. For example, in Figure \ref{fig:4}(b) we have found that vertical velocity along the jet is less sensitive to the increase of resistivity. In Figure \ref{fig:4}(c), we show the temperature along the jet as a function of time for different values of $\kappa$ and $\eta=3.97\times10^{-8}$. In this case, we can see that the temperature increases slightly when the value of the thermal conductivity is higher. In Figure \ref{fig:4}(d), we show that the vertical component of velocity along the jet is higher about time $t\sim 100$ s when the value of $\kappa$ is higher, for all other times the velocity remains without significant variations. In Figures \ref{fig:4}(b) and \ref{fig:4}(d), we can see that the vertical velocity along the jet shows a bump about the time $t\sim 100$ s, this is due to the rapid acceleration of the plasma produced by magnetic reconnection, which is triggered by the magnetic loops close together with opposite polarity and the inclusion of resistivity as indicated in \citet{2.5Dspicules}. In Figure \ref{fig:4}(e), we show the average of jet-apex temperature as a function of $\eta$ for $\kappa=0$ at time $t=210$ s. For the values of $\eta$ used, the temperature increases by 2.1\%. Finally, in Figure \ref{fig:4}(f), we show the average of jet-apex temperature as a function of $\kappa$ and $\eta=3.97\times10^{-8}$. In this case we can see that temperature of the jet-apex increases by $\sim 8.1\%$ with respect to the temperature obtained with the smallest value of conductivity.

% --------------------------------------------------------------
% ----->        Conclusions and Final comments     <-----
% --------------------------------------------------------------

\section{Conclusions and final comments}
\label{sec:conclusions}

In this paper we perform numerical simulations to explore the effects of magnetic resistivity and thermal conductivity in the dynamics and parameters of jet structures with some characteristics of Type II spicules. This paper is a continuation of the work presented in \cite{2.5Dspicules}, where we show that jets with features of Type II spicules and cool coronal jets can be formed as a result of magnetic reconnection in a scenario with magnetic resistivity.

According to the analysis carried out in this paper, we found that the inclusion of thermal conductivity along the magnetic field lines affects the morphology of the jets, in particular the combination $\kappa=32755$ and $\eta=8\times10^{-8}$ makes the jet structure 0.1 Mm wider at its lower part and 0.28 Mm thinner at its apex, measured with respect to jets where only resistivity is considered. In addition, the increase in thermal conductivity makes the jets reach maximum heights of about 7.7 Mm, such heights are 0.4 Mm larger compared to jets with only resistivity. According to the results of the Runs \#5 to \#8 shown in Table \ref{table1}, we can see that an increase of the order of 10 times the value of thermal conductivity causes the jet to reach a maximum height 0.2 Mm larger, it also makes the jet-apex 0.28 Mm thinner, this means that the inclusion of thermal conductivity makes the jet have a width closer to the value of observations. One of the most important results is that the jet-apex heat up and reach temperatures of the order of 73000 K for the combination $\eta=3.97\times10^{-8}$ and $\kappa=32755$ in comparison with the combination $\eta=3.97\times10^{-8}$ and $\kappa=0$, in which the jet-apex temperature is of the order 58000 K. These results are similar to those obtained in \cite{Kuzma_et_al_2017}, however in this paper the formation and evolution of solar spicules use numerical simulations triggered with a vertical velocity pulse that is launched from the upper chromosphere. Instead in our analysis, we do not perturb the solar atmosphere with any pulse at initial time, rather it is the magnetic reconnection that accelerates the plasma and forms the jet. Another difference is that we use a range of resistivity and thermal conductivity values consistent with a fully ionized solar atmosphere.

We have also found, that the increase in the resistivity does not affect the morphology of the jets, and although it slightly increases the  jet-apex temperature it was found that it does not modify the behavior of the temperature along the whole jet structure.

Note that although the model of two magnetic loops close together with opposite polarity is very simple and approximate, this configuration can support the formation of jets mimicking some properties of Type II spicules and cool coronal jets. Furthermore, the inclusion of resistivity and thermal conductivity is consistent with the physical properties found in the solar atmosphere. In particular, the resistivity supports development of the magnetic reconnection process, and thermal conductivity helps the heat to propagate more efficiently along the magnetic field lines.  

These results are important to understanding of nature of spicules. The reason is that two different values of thermal conductivity produce spicules with different temperature and maximum height. From other side, the problem could be degenerate, because two jets with the same temperature and height  could be obtained with and without thermal conductivity at the price of modifying for example the magnetic field, or the temperature model of the chromosphere-corona interface. Eventually the addition of ingredients to a model-simulation will have to face the observational restrictions that will in turn refine the parameter values of simulations and the degeneracy of the problem can be an important problem in itself. 
Finally, it is important to mention the role of heat transfer prior jet propagation, in this case the time scale of physical process related to the thermal conductivity is much smaller than the time needed for jet development and reach the maximum height. At the beginning of simulation runs, the temperature gradient has it's maximum value at the interface, and at the same time the field lines are nearly parallel to this gradient near the foot points, therefore heat transfer influences on initial numerical background in terms of temperature changes even before the jet propagates up. This is important that deserves special attention.

\acknowledgments
This research is partly supported by the following grants: Newton Fund, Royal Society-Newton Mobility grant NI160149, CIC-UMSNH 4.9, and CONACyT 258726 (Fondo Sectorial de Investigaci\'on para la Eduaci\'on). The simulations were carried out in the facilities of the Big Mamma cluster at the LIASC-IFM and in the cluster of the CESCOM-UNAM. V.F. and G.V. thank the STFC for their financial support, grant No ST/M000826/1. V.F. would like to thank the International Space Science Institute (ISSI) in Bern, Switzerland, for the hospitality provided to the members of the team "The Nature and Physics of Vortex Flows in Solar Plasmas". J.J.G.A. thanks to C\'atedras CONACyT (CONACyT Fellow) for supporting this work. Visualization of the simulations data was done with the use of the VisIt software package \citep{Childs_et_al_2012}. This research has received financial support from the European Union's Horizon 2020 research and innovation program under grant agreement No. 824135 (SOLARNET).

\bibliographystyle{aasjournal}

\end{document}